\documentclass[fleqn,usenatbib,useAMS]{mnras}
\newcommand{\RNum}[1]{\uppercase\expandafter{\romannumeral #1\relax}}
\usepackage{float}
\usepackage{adjustbox}
\usepackage{rotating}
\usepackage{CJKutf8}
\usepackage{tablefootnote}
\usepackage{listings}
\restylefloat{table}
\usepackage[caption = false]{subfig}
\usepackage{graphicx}
\usepackage{amsmath}
\usepackage{colortbl}
\usepackage{booktabs} 
\usepackage{amssymb}
\usepackage{multicol}        
\usepackage{bm}	
\usepackage{pdflscape}

\usepackage[T1]{fontenc}
\usepackage{ae,aecompl}
\usepackage{newtxtext,newtxmath}
\title{200,000 Candidate Very Metal-poor Stars in Gaia DR3 XP Spectra}
\author[Y. Yao et al.]{
Yupeng Yao (\begin{CJK*}{UTF8}{gkai}姚宇鹏\end{CJK*}),$^{1}$
Alexander P. Ji,$^{1,2}$
Sergey E. Koposov,$^{3,4,5}$
Guilherme Limberg$^{1,2,6}$
\\
$^{1}$Department of Astronomy \& Astrophysics, University of Chicago, 5640 S Ellis Avenue, Chicago, IL 60637, USA\\
$^{2}$Kavli Institute for Cosmological Physics, University of Chicago, Chicago, IL 60637, USA\\
$^{3}$Institute for Astronomy, University of Edinburgh, Royal Observatory, Blackford Hill, Edinburgh EH9 3HJ, UK\\
$^{4}$Institute of Astronomy, University of Cambridge, Madingley Road, Cambridge CB3 0HA, UK\\
$^{5}$Kavli Institute for Cosmology, University of Cambridge, Madingley Road, Cambridge CB3 0HA, UK\\
$^{6}$Universidade de S\~ao Paulo, IAG, Departamento de Astronomia, Rua do Mat\~ao 1226, Cidade Universit\'aria, SP 05508-090, S\~ao Paulo, Brasil
}

\begin{document}
\label{firstpage}
\pagerange{\pageref{firstpage}--\pageref{lastpage}}
\maketitle

\begin{abstract}
Very metal-poor stars ($\rm[Fe/H] < -2$) in the Milky Way are fossil records of early chemical evolution and the assembly and structure of the Galaxy. However, they are rare and hard to find. Gaia DR3 has provided over 200 million low-resolution ($R \approx 50$) XP spectra, which provides an opportunity to greatly increase the number of candidate metal-poor stars. In this work, we utilise the \texttt{XGBoost} classification algorithm to identify $\sim$200,000 very metal-poor star candidates. Compared to past work, we increase the candidate metal-poor sample by about an order of magnitude, with comparable or better purity than past studies. Firstly, we develop three classifiers for bright stars ($BP$ $<$ 16). They are Classifier-T (for Turn-off stars), Classifier-GC (for Giant stars with high completeness), and Classifier-GP (for Giant stars with high purity) with expected purity of 52\%/45\%/76\% and completeness of 32\%/93\%/66\% respectively. These three classifiers obtained a total of 11,000/111,000/44,000 bright metal-poor candidates. We apply model-T and model-GP on faint stars ($BP$ $>$ 16) and obtain 38,000/41,000 additional metal-poor candidates with purity 29\%/52\%, respectively. We make our metal-poor star catalogs publicly available, for further exploration of the metal-poor Milky Way.
\end{abstract}


\section{INTRODUCTION}
Very metal-poor stars (VMP,$\rm[Fe/H]$$< -2$\footnote{Standard nomenclature would be Very Metal-Poor for $\rm[Fe/H] < -2$. From here we will refer to very metal-poor as just metal-poor.}; \citealt{beers2005discovery}) are fossil records of early chemical enrichment history of the universe. The most metal-poor stars are likely to be some of the oldest stars that exist today, and their atmospheres contain information about the abundance pattern of gas in the early universe \citep[e.g.,][]{frebel2015near}. Chemical abundances of a large sample of metal-poor stars can advance our understanding of early nucleosynthesis and thus constrain the early stellar masses, rotation rates, mixing processes, explosion energies, compact remnant masses (neutron stars or black holes), thermohaline convection and other stellar properties \citep[e.g.][]{heger2010nucleosynthesis, limongi2012presupernova, wanajo2018physical, jones2019new, ishigaki2021origin}. Moreover, chemical abundances for these stars, together with kinematic data, can be utilised to understand the accretion history, and early formation of the Milky-Way \citep[e.g.][see \citealt{helmi2020streams} for a review]{hawkins2015using, das2020ages, horta2021evidence, conroy2022birth, belokurov2022dawn, rix2022poor}.

However, metal-poor stars are rare and difficult to find. Metal-poor stars only make up $\sim$0.1\% of Milky Way stars \citep[e.g.][]{starkenburg2016oldest, el2018most}, and only few thousands of metal-poor stars have been spectroscopically confirmed in past surveys \citep[e.g.][]{placco2018spectroscopic,li2018catalog,chiti2021stellar}. The typical method to search for metal-poor stars is first finding metal-poor candidates and then following up these stars with medium/high-resolution spectra to get more detailed information \citep[e.g.][]{beers2005discovery}. Objective-prism surveys, photometric surveys and some wide area spectroscopic surveys are the major ways to search for metal-poor stars. Objective-prism surveys \citep{bond1970search,bidelman1973brighter,bond1980extremely} were once the most effective method to search for candidate metal-poor stars, which utilised low-resolution spectra ($R \approx 400$) to estimate the strength of the Ca\RNum{2} K line at 393.36 nm. The HK-\RNum{1}, HK-\RNum{2}, and Hamburg/ESO surveys \citep{beers1985search,beers1992search,frebel2006bright,christlieb2008stellar,beers2017bright} found a total of $\sim$4500 VMP stars \citep{limberg2021dynamically}. More recently, photometric surveys are utilised to identify candidate metal-poor stars. SkyMapper Southern Sky Survey (SMSS) utilises SkyMapper $v$ filter that reflect Ca\RNum{2} H\&K absorption features, together with SkyMapper $u, g, i$ photometry to derive metallicities \citep{onken2019skymapper, chiti2021stellar}. Analogously, Pristine utilises a narrow-band filter that is centred on the Ca\RNum{2} H\&K absorption lines, combined with SDSS broad-band $g$ and $i$ photometry to derive metallicities \citep{starkenburg2017pristine,aguado2019pristine}. Javalambre Photometric Local Universe Survey (J-PLUS) \citep{cenarro2019j} and the Southern Photometric Local Universe Survey (S-PLUS) \citep{mendes2019southern} are also photometric surveys which utilise four SDSS-like ($g, r, i, z$) and one modified SDSS ($u$), and seven narrow-band filters to identify low-metallicity stars in the Galactic halo \citep{placco2021splus,placco2022mining,galarza2022j}. Another photometric selection method is Best \& Brightest \citep{schlaufman2014best} which utilises all-sky APASS optical, 2MASS near-infrared, and WISE mid-infrared photometry to identify bright metal-poor star candidates through their lack of molecular absorption near 4.6 microns \citep{placco2019r, reggiani2020most, limberg2021targeting}. Besides the aforementioned dedicated efforts, there are some large surveys that directly observe samples of stars at intermediate resolution spectra and estimate their metallicity, e.g., SEGUE, LAMOST, and RAVE surveys. These surveys have found several thousand of metal-poor stars. The Sloan Digital Sky Survey \citep[SDSS;][]{eisenstein2011sdss}, and its Sloan Extension for Galactic Understanding and Exploration \citep[SEGUE;][]{yanny2009segue} survey ($R \approx 2,000$), SEGUE-1 and -2, which motivated several high-resolution follow-up campaigns \citep[e.g.,][]{aoki2012high}. The Large Sky Area Multi-Object Fiber Spectroscopic Telescope (LAMOST) survey \citep[$R \approx 1800$;][]{deng2012lamost}, which has also triggered some high resolution observations \citep[e.g.,][]{li2022four}. LAMOST-\RNum{1}(DR7) released more than seven million spectra of stars in the Milky Way. The RAdial Velocity Experiment (RAVE; $R \approx 7,000$) \citep{kunder2017radial} delivered spectra for about 480,000 stars. However, the number of candidate metal-poor stars found from each survey is about a few dozens to at most a few thousand, which is too small for a statistical investigation on metal-poor stars, especially for extremely metal-poor ($\rm[Fe/H] < -3$) or ultra metal-poor regime ($\rm[Fe/H] < -4$). Thus we need a survey that can provide a much larger number of stellar spectra to enable us to find such objects.

The \textit{Gaia} mission has brought a revolutionary change to Milky Way astronomy, because it provides astrometric data for billions of stars \citep{collaboration2016gaia, collaboration2022gaia}. In Gaia Data Release 3 (DR3), it released 200 million low-resolution XP spectra \citep[$R \approx 50$;][]{de2022gaia}. Because of its low-resolution, the XP spectra can't provide detailed element abundances of stars. Additionally, Gaia GSP-Phot also does not provide accurate metallicity estimations for the most metal-poor stars \citep[]{andrae2022gaia}. However, some works have demonstrated that these low resolution XP spectra can be utilised to estimate effective temperature, surface gravity, and metallicity \citep[e.g.,][]{xylakis2022method, andrae2023robust,zhang2023parameters}. Thus, these 200 million low-resolution XP spectra give us an opportunity to greatly increase the number of candidate metal-poor stars, if we can make full use of them. 

In this work, we identify metal-poor stars in the Gaia DR3 XP spectra using the \texttt{XGBoost} classification algorithm. In Section \ref{data}, we describe the XP spectra and other data we utilised in this work. In Section \ref{Methodology}, we introduce \texttt{XGBoost}, discuss the training process, and evaluate the performance of the models. Then, we utilise \texttt{XGBoost} models to make a prediction on the XP spectra, shown and discussed in Section \ref{Results}. Then, we compare our work with other surveys and projects and utilise existing high-resolution spectroscopic data to validate the performance of our models in section \ref{Discussion}. Finally, we summarize this work in \ref{summary}.

\section{DATA}\label{data}

\begin{figure*}
    \centering
    \includegraphics[scale=0.30]{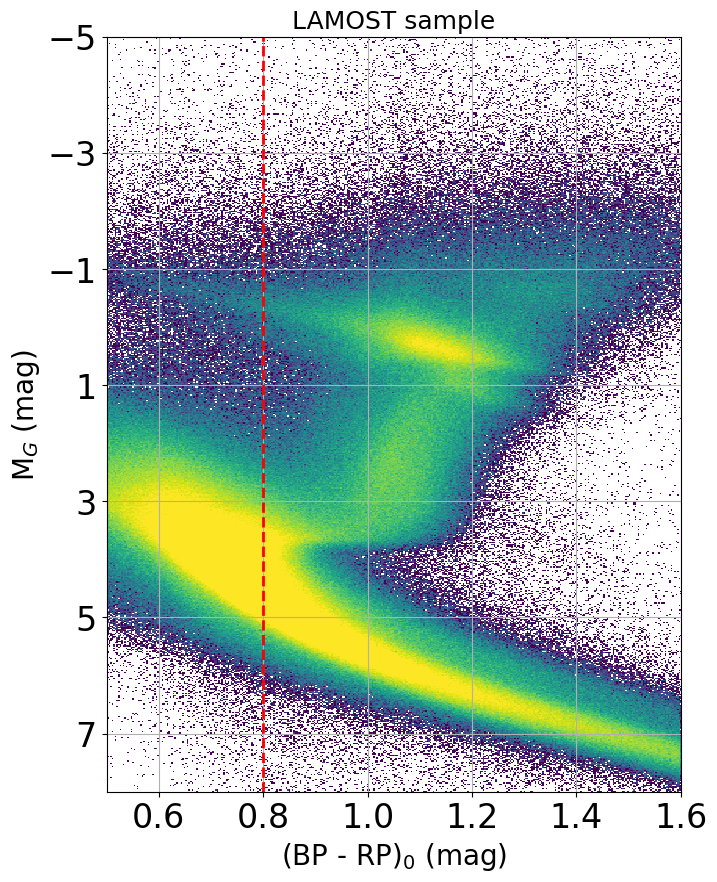}
    \includegraphics[scale=0.30]{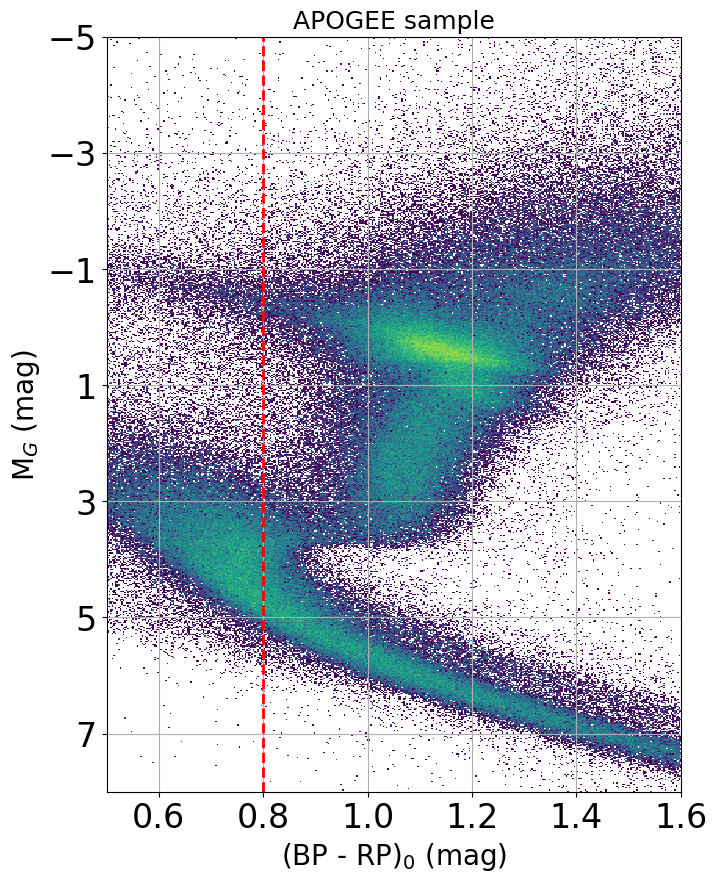}
    \includegraphics[scale=0.30]{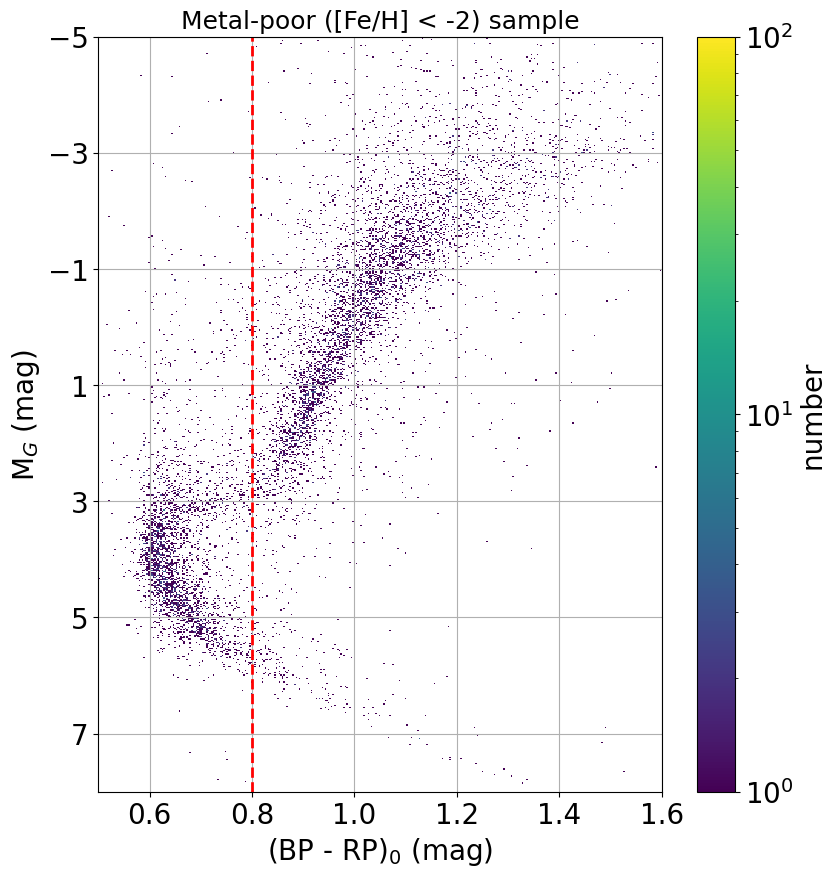}
    \caption[frog]{Colour-magnitude diagram of our training and testing sets. The horizontal axis is the Gaia intrinsic colour $(BP-RP)_{0}$, the vertical axis is the Gaia absolute $G$ magnitude. The LAMOST and APOGEE samples, which primarily comprise main-sequence turn-off, giants, and dwarfs stars, are shown in the left and middle panels. The right panel shows the metal-poor stars from LAMOST and APOGEE. Metal-poor stars are primarily turn-off and giant stars. We divide the training and testing set into two parts, according to $(BP-RP)_{0}$, as shown in the red dashed line in the Figure. On the left/right side of red dashed line are the samples utilised to train the model to identify the turn-off/giants metal-poor stars.}
    \label{fig:npr_distribution}
\end{figure*}

\subsection{Data Sets}

In this work, the data utilised include Gaia DR3 XP spectra \citep{de2023gaia}, Gaia DR3 photometry \citep{vallenari2023gaia}, LAMOST DR7 \citep{LAMOST} metallicity and Apache Point Observatory Galactic Evolution Experiment (APOGEE; \citealt{apogee2017}) DR17 \citep{abdurro2022seventeenth} metallicity.

\textbf{Gaia XP spectra:} Gaia DR3 released low-resolution blue and red photometer spectra ($BP$/$RP$ or XP spectra) for 210 million stars. Metallicities were derived from these spectra in the Gaia GSP-Phot, but they are not accurate at low metallicities\citep{andrae2022gaia}. Thus, it is not efficient to directly utilise the GSP-Phot metallicity [M/H] in Gaia DR3 to search for metal-poor stars. The XP spectra have wide wavelength coverage (330 to 1050nm) and low-resolution. Because of its wide wavelength coverage, strong lines valuable for metallicity estimation are covered in it, such as Ca \RNum{2} K and Ca \RNum{2} infrared triplet, as well as broad-band or narrow-band photometry. Thus, in theory, XP spectra can be utilised to detect metal-poor stars. The XP spectra are released as Hermite function coefficients rather than fluxes v.s. wavelength \citep{carrasco2021internal}. In order to avoid information loss \citep{carrasco2021internal}, the input for \texttt{XGBoost} model are XP spectra coefficients, rather than corresponding sampled XP spectra. \texttt{XGBoost} requires the input vectors to be of the same length, so we do not truncate the XP coefficients. 

Before inputting the XP coefficients to the model,  we first normalize and deredden them. We normalized XP coefficients by their first coefficient to remove apparent magnitude information. Additionally, to take into account reddening, we determined the extinction coefficients  $\pmb \alpha$, $\pmb \beta$, $\pmb \gamma$ to correct the normalized XP coefficient vectors $\textbf C$ for extinction ${\textbf C}_{corrected} = {\textbf C} -  ({\pmb \alpha} + {\pmb \beta} \hat{\textbf {C}} )E_{B-V} - {\pmb \gamma} E_{B-V}^2$ . Here, the $\hat{\textbf {C}}$ is a truncated XP coefficient vector with first 10 elements, $\pmb \alpha$, $\pmb \gamma$ are vectors and $\pmb \beta$ is a matrix. We fit for $\pmb \alpha$, $\pmb \beta$, $\pmb \gamma$ by taking high extinction stars in APOGEE and matching them with stars with  similar $\log g$ (surface gravity), $T_{\rm eff}$ (effictive temperature) and metallicity, but at low extinction. The extinction utilised in this analysis is from a 2-D map by \citet{schlegel1998maps}. 

\textbf{Gaia DR3 photometry:} We also utilised Gaia DR3 photometry\citep{vallenari2023gaia} in this work. Gaia’s $G$ band covers a wavelength range from near ultraviolet ($\sim$330 nm) to the infrared ($\sim$1050 nm). The other two bands, denoted $BP$ and $RP$, cover smaller wavelength ranges, from approximately 330 to 680 nm, and 630 to 1050 nm respectively\footnote{\url{https://www.cosmos.esa.int/web/gaia/edr3-passbands}}. We utilise the extinction law, as described in \url{https://www.cosmos.esa.int/web/gaia/edr3-extinction-law}, to get the intrinsic colour $(BP - RP)_0$.

\textbf{LAMOST DR7 and APOGEE DR17 metallicity:} In order to train our model to identify metal-poor stars, we need a sample of stars that already have reliable metallicity estimates to provide true labels. We utilised the spectroscopic metallicity from the LAMOST DR7 \footnote{\url{https://dr7.lamost.org/}} and APOGEE DR17 \footnote{\url{https://www.sdss4.org/dr17/}}. LAMOST spectra ($R\approx 1800$) cover the optical band from 370 to 900 nm. APOGEE spectra ($R \approx 22,500 $) are a good complement to LAMOST, because they cover the infrared band from 1.51 to 1.70 ${\mu}$m, which is more suited for dust extincted regions, i.e., the Galactic disk and bulge. In total, we have $4\times10^{6}$ LAMOST and $6.5\times10^{5}$ APOGEE stars.

Data Queries and Quality Cuts: We utilised the Whole Sky Database (WSDB)\footnote{\url{https://www.ast.cam.ac.uk/ioa/wikis/WSDB/index.php/Main_Page}} for all queries (see Appendix~\ref{queries} for the ADQL queries), which ingested the entire catalog for APOGEE DR17 and LAMOST DR7. We did not do any significant quality cuts, but we do not think this will significantly affect the results for a few reasons. First, classification models are less sensitive to quality cuts than regression models. Second, after comparing the overlapping very metal-poor stars in LAMOST and APOGEE, we found that even if a star is flagged as bad spectral fitting solutions in either APOGEE or LAMOST, it often still carries sufficient information regarding being very metal-poor or not. For example, for LAMOST we adopted quality flags of SNR $> 20$ and feh\_err $<0.5$ \citep[e.g.,][]{zhang2023parameters}, which removed 22\% of our metal-poor training set. However, overlapping APOGEE spectra suggested that 84\% of these were actually still very metal-poor. For APOGEE, metal-poor stars run up against the edge of the spectral grid, so using quality flags (e.g., FE\_H\_FLAG = 0) removed all stars with $\mbox{[Fe/H]} < -2.25$ even though they are very metal-poor in LAMOST.

\subsection{Training and testing sets}
The Gaia XP spectra with the LAMOST or APOGEE metallicity form the training and testing set in this work. We directly put them together because the average difference between LAMOST and APOGEE$\rm[Fe/H]$is 0.007 dex, which is well below the typical uncertainty in metallicity of LAMOST ($>$ 0.2 dex) or APOGEE ($>$ 0.1 dex). Therefore, we conclude that these surveys are on similar metallicity scales within the range of parameters tested. Before the training process, we need to set some constraints on the training and testing set by intrinsic colour $(BP - RP)_0$, magnitude $BP$ and extinction $E(B-V)$.

For the training set, we only consider stars with $(BP - RP)_0 > 0.5$, because, as shown in the right panel of Figure \ref{fig:npr_distribution}, we do not have metal-poor samples with $(BP - RP)_0 < 0.5$. Note that the method utilised to calculate the $(BP - RP)_0$ excludes almost all of the $E(B-V)$ > 2 stars, because the extinction coefficients should not be extrapolated outside the extinction range of this algorithm, as described in \url{https://www.cosmos.esa.int/web/gaia/edr3-extinction-law}. Additionally, we exclude fainter stars ($BP$ > 16) in the training set, because the XP spectra with $BP$ > 16 generally do not have high signal-to-noise ratio (S/N < 300). Thus, for the training set, we only consider stars that satisfy the following criteria:

(\RNum{1}) 0.5 < $(BP - RP)_0$ < 1.6

(\RNum{2}) $E(B-V)$ < 2

(\RNum{3}) $BP$ < 16

However, for the testing set, we only constrain the data by 0.5 < $(BP - RP)_0$ < 1.6 and $E(B-V)$ < 2. We aim to see whether our classifiers that are trained on bright stars ($BP$ $< 16$) can be utilised to identify the faint metal-poor stars ($BP$ $> 16$). Thus, we include stars that satisfy the following criteria in the testing set:

(\RNum{1}) 0.5 < $(BP - RP)_0$ < 1.6

(\RNum{2}) $E(B-V)$ < 2

After applying cuts, we get $2.5\times10^{6}$ LAMOST stars and $4.5\times10^{5}$ APOGEE stars with XP spectra available, of which 4088 and 1295, respectively, are metal-poor stars with $\rm[Fe/H] < -2$. In total, we utilise $2.9\times10^{6}$ spectra for training and testing, of which 0.2\% are metal-poor stars. We select $4\times10^{5}$ of them as testing set and $2.5\times10^{6}$ of them as training set. 

Figure \ref{fig:npr_distribution} shows the colour-magnitude diagram of our training and testing sample. The horizontal axis is intrinsic  colour $(BP-RP)_{0}$, the vertical axis is the absolute $G$ magnitude (without any parallax cut here). The left and middle panels show the stars from the LAMOST and APOGEE surveys in the training and testing sets, which comprises main-sequence turn-off, dwarf, and giant stars. The right panel shows the distribution of metal-poor stars in the training and testing set. The majority of metal-poor stars are turn-off and giant stars. Figure \ref{fig:npr_distribution} suggests that our algorithm should only confidently identify metal-poor giants and turn-off stars, because metal-poor stars in other evolutionary stages would be extrapolation. Note that it is harder to find very metal-poor turn-off stars than giants. Because the resolution of XP spectra are low, the information we can get from them are close to what we can get from narrow band photometric surveys, but the photometric features of turn-off stars are less metallicity dependant, because they are hotter and absorption features are suppressed. Consequently, we utilise different models to find metal-poor turn-off and giant stars. We divide the training and testing sample into two parts, according to $(BP-RP)_{0}$ <  0.8 or >  0.8. The models trained on the former dataset are responsible for finding turn-off metal-poor stars, and the other models trained on the latter dataset are in charge of the giant metal-poor stars. As shown in the right panel of Figure \ref{fig:npr_distribution}, our dataset does not have many metal-poor dwarf stars, so we do not expect to find low-metallicity dwarf stars in this work. 

\begin{figure*}
    \centering
    \subfloat[]{\includegraphics[scale=0.35]{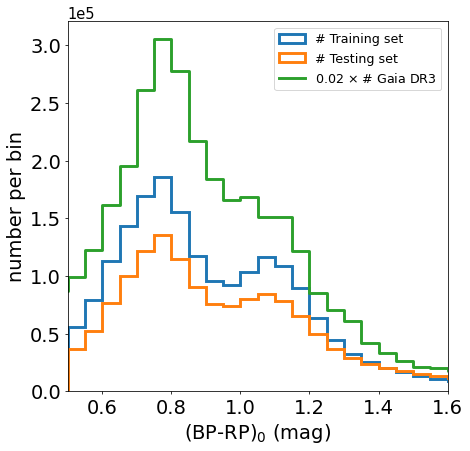}}
    \subfloat[]{\includegraphics[scale=0.35]{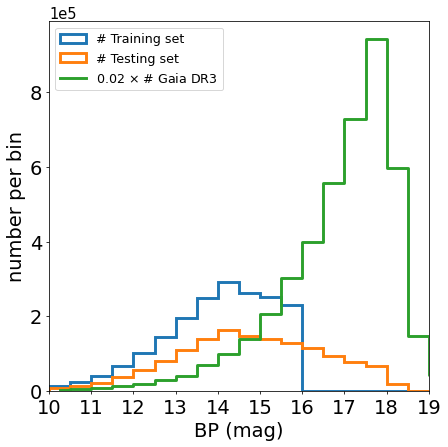}}
    \subfloat[]{\includegraphics[scale=0.35]{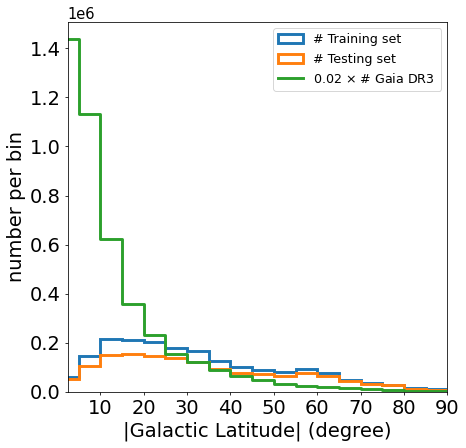}}\\
    \caption{$(BP-RP)_{0}$, $BP$, and absolute Galactic latitude distribution of training, testing and Gaia DR3 with $(BP-RP)_{0} > 0.5$ and $E(B-V) < 2$ in this plot, which have XP spectra. We only randomly select 2\% of the Gaia DR3 data with XP spectra to display in this figure.}
    \label{fig: ttg}
\end{figure*}

The $(BP-RP)_{0}$, $BP$, and $|b|$ distribution of Gaia DR3 data, training and testing set are shown in Figure \ref{fig: ttg}. Note that, we only include Gaia DR3 data with $(BP-RP)_{0} > 0.5$ and $E(B-V) < 2$ that have XP spectra in this plot. We see that the distributions of the Gaia data included in this plot are pretty different from our training and testing set, especially for the $(BP-RP)_{0}$ and Galactic latitude $b$ distributions, which reminds us that the metal-poor candidates we find may only be a small fraction of the total.   

\section{Model training and validation}\label{Methodology}

We choose the \texttt{XGBoost} algorithm to find metal-poor stars because it is a powerful and flexible algorithm that has been utilised in variety of sub-fields of astrophysics \citep[e.g.,][]{li2021identification, he2022identification, pham2022follow, lucey2022over, rix2022poor}. The algorithmic principles for \texttt{XGBoost} are not complex. In short, \texttt{XGBoost} repeatedly builds decision trees to fit the residuals from the previous tree, until the residuals stop shrinking or it reaches the maximum number of trees, which is a free parameter. Then it sums the results from each tree, which are weighted by a learning rate ($\eta$), and plug this value into the Sigmoid function, $\sigma(x) = \frac{1}{1 + e^{-x}}$, to calculate the probability of the input belonging to a certain category. For a detailed description of \texttt{XGBoost}, see \citet{chen2016xgboost}. 

In this work, we utilise the coefficients of normalized and dereddened XP spectra together with their corresponding$\rm[Fe/H]$from LAMOST or APOGEE to compose training and testing sets to train the \texttt{XGBoost} model to identity metal-poor stars in Gaia DR3. We describe the training process and the performance of the well-trained models in this section.

\subsection{Training process}
In this work, we choose multi-classification algorithm to identify the metal-poor stars. The metallicity ([Fe/H]) of the training and testing samples ranges from $-$2.5 to $+$1.0. We utilise \texttt{XGBoost} models to classify the stars into four metallicity intervals:$\rm[Fe/H]$< $-$2.0, $-$2.0 <$\rm[Fe/H]$< $-$1.5, $-$1.5 <$\rm[Fe/H]$< $-$1.0, and $-$1.0 <$\rm[Fe/H]$< $+$1.0, with probabilities $P_{0}, P_{1}, P_{2}, P_{3}$ respectively. For a star, when its $P_{0}$ is larger than the other probabilities, it will be classify as metal-poor star. The prediction uncertainty can be calculated from the probabilities of the multi-classification result, see appendix \ref{Appendix_a} for more details. We choose the \texttt{XGBoost} classification algorithm, rather than the regression algorithm, for following four reasons. (\RNum{1}) The minimum$\rm[Fe/H]$of the training and testing set is $-$2.5, because of LAMOST and APOGEE analyses limitations, even though we do know there exist metal-poor stars with$\rm[Fe/H]$< $-$2.5 in the data set. (\RNum{2}) Regression would waste a lot of computational power on deciding the specific metallicity value for non-metal-poor stars ([Fe/H] > $-$2.0) which we do not care about. (\RNum{3}) Unlike a regression algorithm, classification algorithm can more easily trade off completeness against purity. For samples that are difficult to identify, for example, turn-off stars and faint stars, we can sacrifice completeness for higher purity.  

We utilise completeness and purity calculated on the test set to evaluate the performance of the models. Completeness refers to how completely our model can find all of the metal-poor stars. Purity refers to the fraction of true metal-poor stars for the set predicted to be metal-poor by our models. Completeness and purity are defined as:
\begin{equation}
\text{Completeness} = \frac{\text{True\;positive}}{\text{True\;positive + False\;negative}}
\label{eq:completeness}
\end{equation}

\begin{equation}
\text{Purity} = \frac{\text{True\;positive}}{\text{True\;positive + False\;positive}}
\label{eq:purity}
\end{equation}
Positive and negative samples here refer to the metal-poor ([Fe/H] $<-2$) and non-metal-poor ([Fe/H] $>-2$) stars respectively. We divide the input samples into two training sets, according to their intrinsic colour: 0.5 < $(BP-RP)_{0}$ < 0.8 and 0.8 < $(BP-RP)_{0}$, as shown in Figure \ref{fig:npr_distribution}, to find metal-poor turn-off and giant stars, respectively. Metal-poor giant stars make up 0.26\% of the training set with 0.8 < $(BP-RP)_{0}$. However, metal-poor turn-off stars are much rarer, only make up 0.06\% of the training set with 0.5 < $(BP-RP)_{0}$ < 0.8. Thus, it could be expected that metal-poor turn-off stars will be more difficult to find than metal-poor giant stars.

In preliminary tests, we found that the extreme imbalance between positive ([Fe/H] $< -2$) and negative ([Fe/H] $> -2$) samples badly hinders our training process. To solve this problem, we processed the training sets in the following two steps:

\hangindent 2em
\hangafter=0
Step \RNum{1}: Utilise random under-sampling to randomly remove over-represented metal-rich stars in the training set. The negative ([Fe/H] $> -2$) to positive ([Fe/H] $< -2$) ratio of the training set after under-sampling is defined as NPR. We will change the NPR of the training set from 1 to the maximum value that the training set allowed.

\hangindent 2em
\hangafter=0
Step \RNum{2}: Adopt over-sampling algorithm Synthetic Minority Over-sampling Technique (SMOTE) to populate the metal-poor stars in the training set that has been under sampled. The SMOTE algorithm is an over-sampling method which synthesizes new examples from the minority class by selecting  neighboring examples in the feature space and then synthesizing a new sample at the point along the line connecting these two samples \citep{chawla2002smote}. 

We utilise \texttt{RandomSearchCV} from \texttt{scikit-learn} \citep{pedregosa2011scikit} to tune the \texttt{XGBoost} hyper-parameters. When training \texttt{XGBoost}, a lot of hyper-parameters can be adjusted, such as the learning rate ($\eta$), the maximum depth of a tree, and the minimum loss reduction required to make a further partition on a leaf node of the tree ($\gamma$). In order to find the optimal set of parameters, we utilise \texttt{RandomSearchCV} from \texttt{scikit-learn} \citep{pedregosa2011scikit}. \texttt{RandomSearchCV} will go through points that are randomly selected from the predefined box in hyper-parameter space, as shown in below, to find the optimal set of parameters.

\begin{itemize}
  \item $\rm n\underline{~~}estimators$: from 100 to 1200 in steps of 50
  \item $\rm max\underline{~~}depth$: from 2 to 15 in steps of 1
  \item $\rm learning\underline{~~}rate$: from 0.05 to 1 in steps of 0.05
  \item $\rm subsample$: from 0.5 to 1 in steps of 0.05
  \item $\rm colsample\underline{~~}bytree$: from 0.3 to 0.9 in steps of 0.05
  \item $\rm min\underline{~~}child\underline{~~}weight$: from 1 to 20 in steps of 1
  \item $\rm gamma$: from 0 to 0.7 in steps of 0.02
\end{itemize}

\begin{figure}
    \centering
    \includegraphics[scale=0.5]{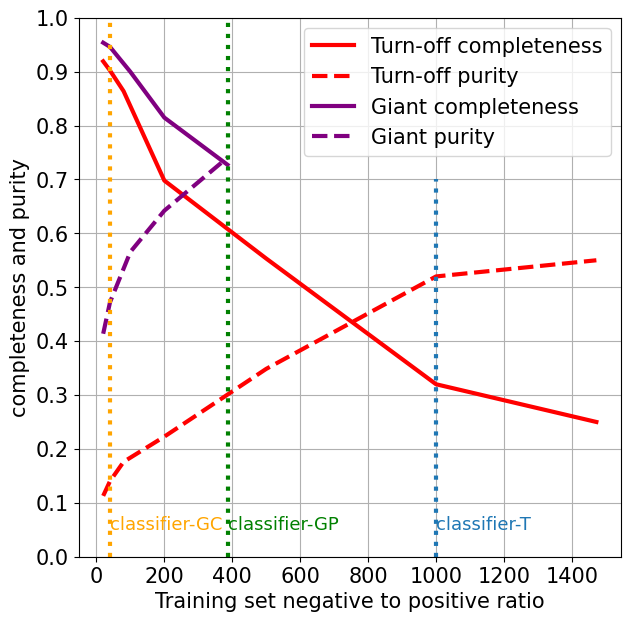}
    \caption{The completeness and purity of classifiers as a function of training NPR. The horizontal-axis is the negative to positive ratio of the training sample; the vertical-axis is completeness and purity of models. At each NPR, the classifiers were ran with the optimized set of hyper-parameters. The vertical lines with different colours refer to the NPR were chosen for Classifier-GC, Classifier-GP, and Classifier-T. The corresponding completeness and purity can be read from the vertical lines. The purple curves refer to the classifiers trained to find metal-poor Giants stars and the red curves refer to the classifiers aimed to find metal-poor turn-off stars.}
    \label{fig:npr_pc}
\end{figure}

In this work, finding metal-poor stars trades off purity for completeness. For each NPR, we utilise \texttt{RandomSearchCV} to find the optimal set of parameters. Figure \ref{fig:npr_pc} shows the completeness and purity of the well optimized model as a function of the training set NPR. The purple curves refer to the classifiers that are trained to find metal-poor giant stars , and the red curves refer to the classifier to find metal-poor turn-off stars. From Fig. \ref{fig:npr_pc} we see that increasing the NPR of the training set will increase the purity but decrease the completeness of the classifiers, and it is much easier to find metal-poor giant stars than metal-poor turn-off stars, just as we discussed before. The three vertical lines indicate the NPR that are chose for Classifier-GP (Green, 386), Classifier-GC (Yellow, 40), Classifier-T (Blue, 1000). Classifier-GC (Giant Complete) here denotes the model utilised to find metal-poor giants with high completeness, Classifier-GP (Giant Pure) denotes the model utilised to find metal-poor giants with high purity, and Classifier-T (Turn-off) denotes the model utilised to find turn-off metal-poor stars. The (completeness, purity) for our Classifier-T, Classifier-GC, Classifier-GP are (40.0\%, 47.2\%), (94.6\%, 47.2\%), (72.7\%, 74.1\%) respectively, which are derived by 3-fold cross-validation.

\subsection{Models evaluation}
After the training process, we utilise the testing sets to evaluate the performance of the classifiers on different [Fe/H], $BP$, $(BP-RP)_{0}$, and absolute Galactic latitude $|b|$. Typically, there are three factors that effect the performance of the classifiers: stellar species (turn-off or giants stars), brightness, and reddening. In this work, we utilise intrinsic colour $(BP-RP)_{0}$ to denote the type of stars, because we do not have metal-poor dwarf stars in the training and testing sets, as shown in Figure \ref{fig:npr_distribution}. $BP$ magnitude denotes the brightness of the stars. Additionally, the absolute $|b|$ can be utilised as an indicator of reddening, because stars in low $|b|$ regions, such as disk and bulge, often have severe extinction.

\begin{figure*}
    \centering
    \includegraphics[scale=0.48]{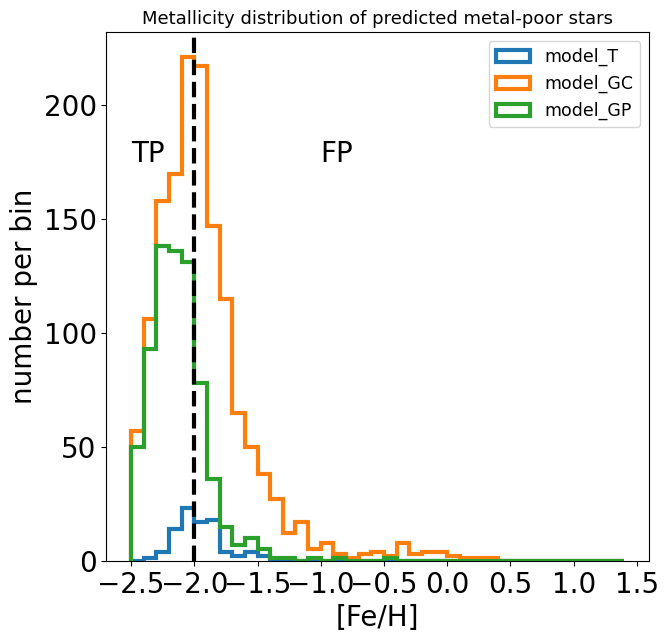}
    \includegraphics[scale=0.48]{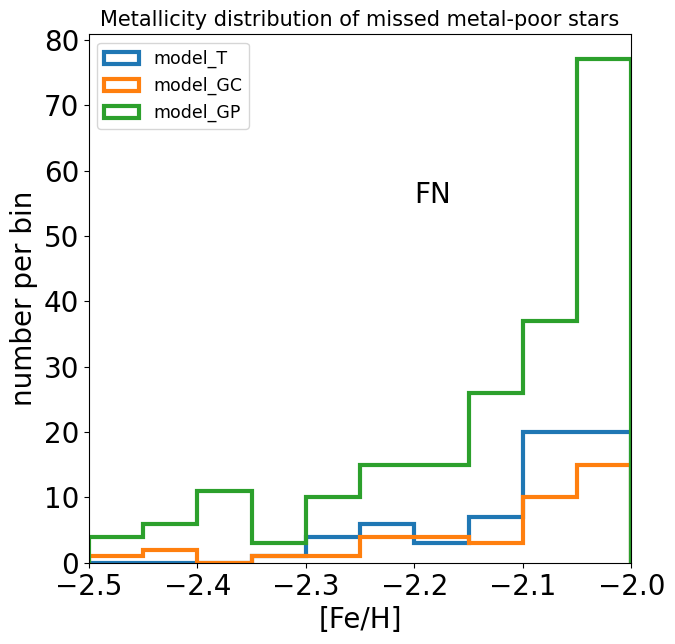}
    \caption{Left panel show the metallicity distribution of stars that are predicted to be metal-poor. The dashed line is the boundary of true positive samples and false positive samples. Right panel shows the metallicity distribution of False-negative stars (i.e. metal-poor stars missed by \texttt{XGBoost}).}
    \label{fig:metallicity_distribution}
\end{figure*}

The metallicity distribution for stars in the testing set classified as metal-poor by different classifiers is shown in Figure \ref{fig:metallicity_distribution}. The metallicity distribution for True Positive (TP), False Positive (FP) and False Negative (FN) samples in the testing set are shown in left and right panels, respectively. Comparing the distributions of Classifier-GC and Classifier-GP in the left panel, we see that the Classifier-GP can effectively remove the FP stars, although it loses some TP stars. On the other hand, the right panel shows that Classifier-GP loses some metal-poor stars with$\rm[Fe/H]$$<$ -2.8, which is the cost of high purity. This is why we provide Classifier-GC as supplement to Classifier-GP. Classifier-GC provides a high completeness dataset and Classifier-GP provide a high purity dataset. The good news for Classifier-GC is that most of the misclassified metal-poor still have rather low metallicity close to the $\rm$\rm[Fe/H]$= -2$ boundary. 

\begin{figure*}
    \centering
    \subfloat[]{\includegraphics[scale=0.27]{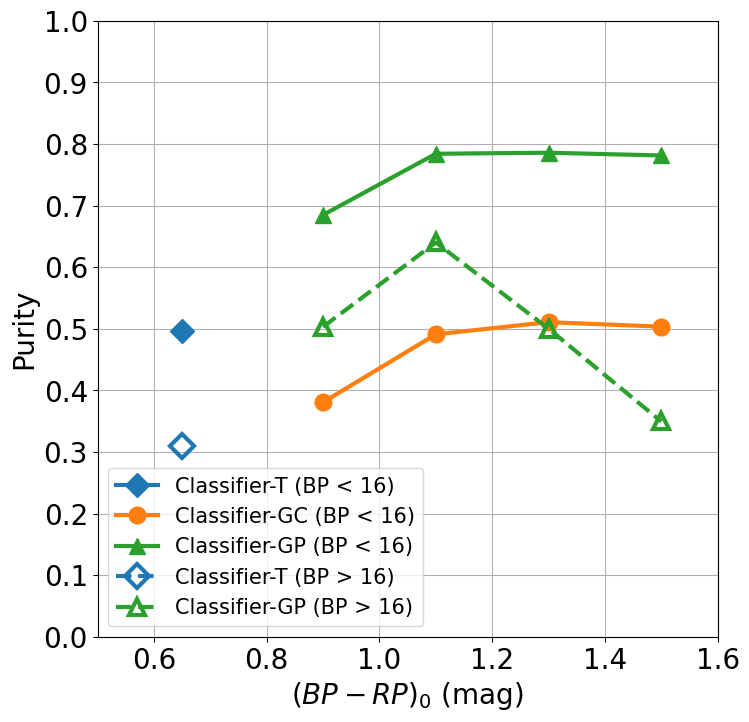}}
    \subfloat[]{\includegraphics[scale=0.27]{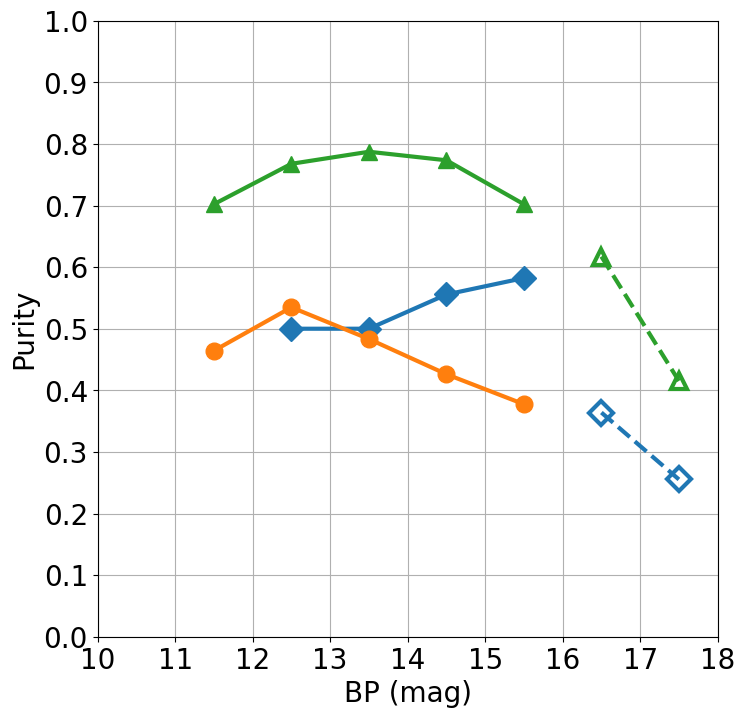}}
    \subfloat[]{\includegraphics[scale=0.27]{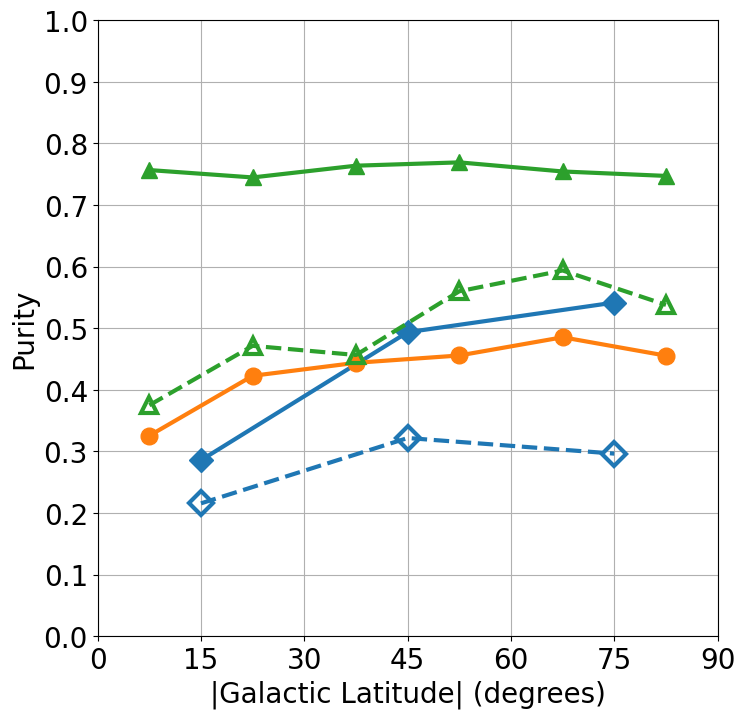}}\\
    \subfloat[]{\includegraphics[scale=0.27]{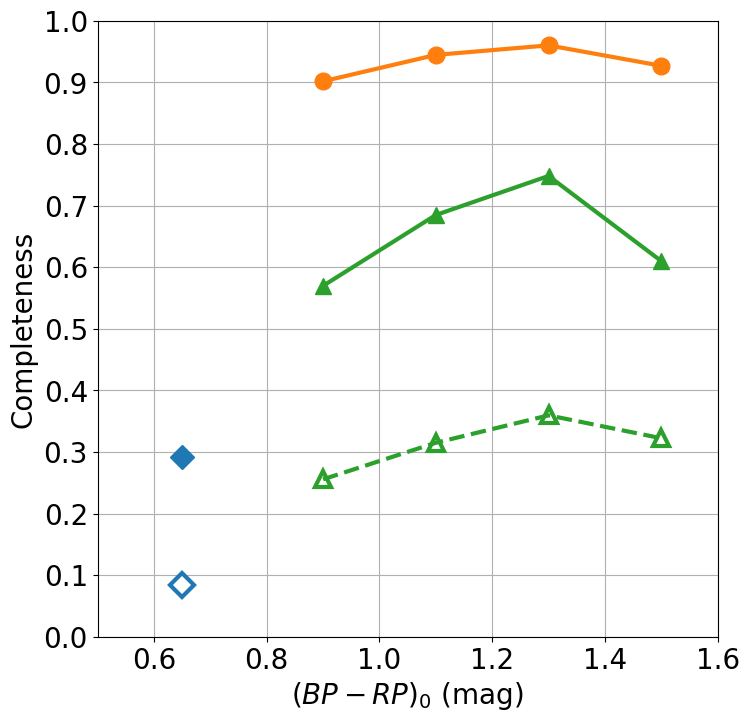}}
    \subfloat[]{\includegraphics[scale=0.27]{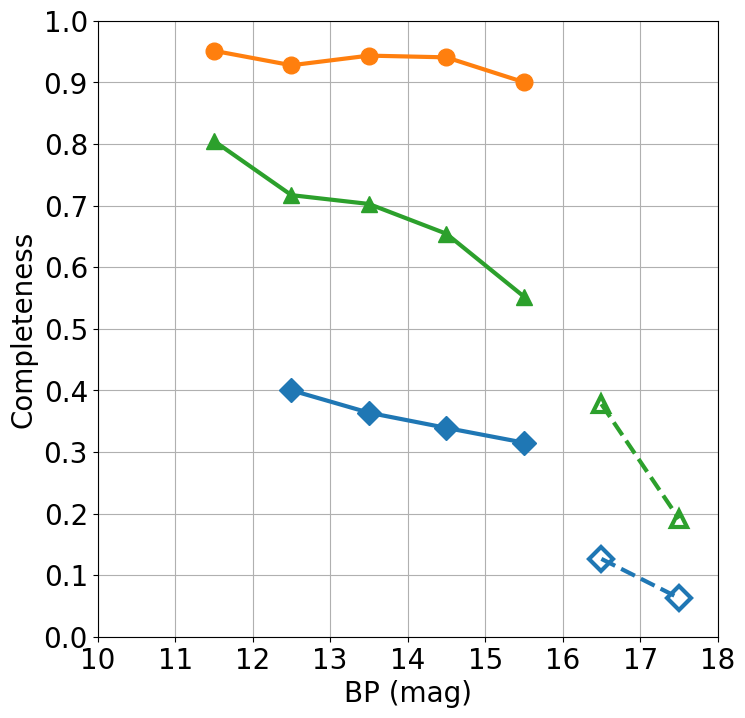}}
    \subfloat[]{\includegraphics[scale=0.27]{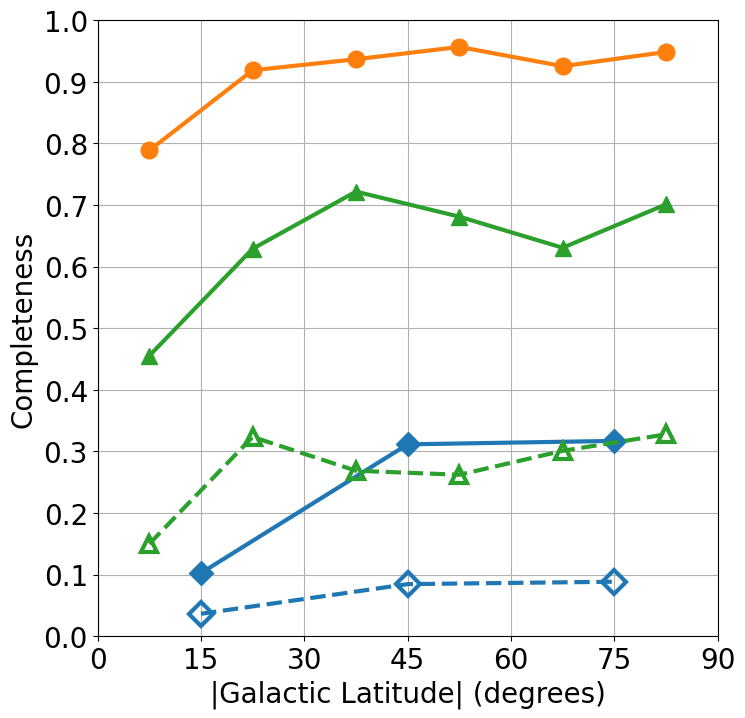}}\\
    \caption{The completeness and purity of different classifiers as a function of intrinsic colour $(BP-RP)_{0}$, $BP$ band magnitude $BP$ and absolute Galactic latitude $|b|$.}
    \label{fig: CP_dis}
\end{figure*}

The completeness and purity distributions of the classifiers on different $(BP-RP)_{0}$, $BP$, and $|b|$ intervals are shown in Figure \ref{fig: CP_dis}. We utilise different colours and symbols to denote different models, and dashed and solid lines to denote faint or bright stars. Let's discuss the performance of the classifiers on bright stars ($BP$ < 16) first. Panel (a) and (d) show the performance of the classifiers as a function of $(BP-RP)_{0}$. We see that Classifier-T has a comparable purity at the blue end of the classifiers-GP and classifiers-GC, but its completeness is lower than these two models, because it is harder to find metal-poor turn-off stars, we have to sacrifice the completeness for high purity, just as we discussed before. Panel (b) and (e) show the performance of classifiers as a function of brightness. We can see that bright stars tend to have higher purity and completeness than faint stars, because bright stars typically have higher signal to noise ratio. Panel (c) and (f) show the performance as a function of $|b|$. The completeness and purity of our classifiers are lower in low-latitude region, because in this region extinction makes classification more difficult even with our coefficients extinction calibrations and higher contamination rate of metal-rich ($\rm [Fe/H] > -2$) stars decrease the purity statistically. Note that, because there are few metal-poor turn-off stars at low or high galactic latitude in our training and testing sets, we increased the bin size for turn-off stars in these two panels to avoid statistical fluctuations.

Most of the stars with XP spectra released by Gaia DR3 are faint ($BP$ > 16), so it is worthwhile to evaluate the performance of the classifiers, which are trained on bright stars, on the faint stars. We utilise Classifier-T and Classifier-GP to make the prediction on faint stars. As shown in the dashed lines and open symbols of Figure \ref{fig: CP_dis}, the overall purity for Classifier-T is 29\%, for Classifier-GP is 52\%. This purity is better than we expected, so we include the faint stars in our catalog. However, as shown in panel (d), (e), (f), the completeness for faint turn-off candidates is pretty low, less than 10\%, which means that the faint metal-poor turn-off stars we have in our final catalogs only make up a very small fraction of the total. Because of the low $S/N$ ratio for faint stars, it is harder for us to find the genuine metal-poor ones. Thus, under this circumstance, purity has a higher priority than completeness. We can make a Shannon-Entropy cut on the final results to increase their purity. More details about the Shannon-Entropy cut are shown in Appendix \ref{Appendix_a}.

\section{Results}\label{Results}
We have three reliable classifiers, Classifier-T, Classifier-GC, and Classifier-GP. We now classify the 200 million XP spectra released in Gaia DR3, and obtain three corresponding candidate metal-poor star catalogs, as shown in Table \ref{table: Classifier_T}, Table \ref{table: Classifier_GC}, and Table \ref{table: Classifier_GP}, which in total contain 200,000 metal-poor candidates. 

\begin{table*}
\begin{tabular}{|p{3cm}|p{1.5cm}|p{1.0cm}|p{1.5cm}|p{1.0cm}|p{0.5cm}|p{0.5cm}|p{0.5cm}|p{0.5cm}|p{2.5cm}|}
 \hline
 \multicolumn{10}{|c|}{Candidates found by Classifier-T} \\
 \hline
Gaia DR3 source id&$(BP-RP)_{0}$&$M_G$&$E(B-V)$&$BP$&$P_{0}$&$P_{1}$&$P_{2}$&$P_{3}$&Shannon Entropy\\
 &(mag)&(mag)&(mag)& & & & & & \\
 \hline
6650038640545499264&0.59&3.66&0.07&15.65&0.97&0.03&0.0&0.0&0.17\\
6650111586271008256&0.58&2.96&0.06&16.34&0.84&0.15&0.0&0.0&0.68\\
6650144949575814016&0.61&-0.24&0.07&17.65&0.83&0.15&0.0&0.02&0.77\\
6650193499886281088&0.62&3.76&0.07&16.65&0.84&0.14&0.01&0.0&0.7\\
6650230470965151360&0.78&3.14&0.08&16.9&0.85&0.1&0.0&0.05&0.76\\
 \hline
\end{tabular}
\vspace*{2.5mm}
\caption{Metal-poor turn-off candidates found by Classifier-T. $P_{0}, P_{1}, P_{2}, P_{3}$ refer to the probability of a stars with $-2.5 < \rm[Fe/H] < -2$, $-2 < \rm[Fe/H] < -$1.5, $-1.5 < \rm[Fe/H] < -1$, $-1 < \rm[Fe/H] < +1$. (This table is available in its entirety in the online supplementary material)}
\label{table: Classifier_T}
\end{table*}

\begin{table*}
\begin{tabular}{|p{3cm}|p{1.5cm}|p{1.0cm}|p{1.5cm}|p{1.0cm}|p{0.5cm}|p{0.5cm}|p{0.5cm}|p{0.5cm}|p{2.5cm}|}
 \hline
 \multicolumn{10}{|c|}{Candidates found by Classifier-GC} \\
 \hline
Gaia DR3 source id&$(BP-RP)_{0}$&$M_G$&$E(B-V)$&$BP$&$P_{0}$&$P_{1}$&$P_{2}$&$P_{3}$&Shannon Entropy\\
 &(mag)&(mag)&(mag)& & & & & & \\
 \hline
4252405961205838208&0.81&-1.55&0.68&15.77&0.6&0.01&0.01&0.38&1.12\\
4252433105401980800&1.5&-8.02&0.61&15.65&0.39&0.21&0.39&0.01&1.57\\
4252454580242134912&0.84&-2.07&0.78&15.1&0.93&0.0&0.02&0.04&0.42\\
6032351905927100928&0.98&-0.61&0.42&15.69&0.56&0.02&0.09&0.34&1.4\\
6032356578851595392&1.13&nan&0.48&15.82&0.95&0.05&0.0&0.0&0.3\\
 \hline
\end{tabular}
\vspace*{2.5mm}
\caption{Metal-poor giant candidates found by Classifier-GC. (This table is available in its entirety in the online supplementary material)}
\label{table: Classifier_GC}
\end{table*}

\begin{table*}
\begin{tabular}{|p{3cm}|p{1.5cm}|p{1.0cm}|p{1.5cm}|p{1.0cm}|p{0.5cm}|p{0.5cm}|p{0.5cm}|p{0.5cm}|p{2.5cm}|}
 \hline
 \multicolumn{10}{|c|}{Candidates found by Classifier-GP} \\
 \hline
Gaia DR3 source id&$(BP-RP)_{0}$&$M_G$&$E(B-V)$&$BP$&$P_{0}$&$P_{1}$&$P_{2}$&$P_{3}$&Shannon Entropy\\
 &(mag)&(mag)&(mag)& & & & & & \\
 \hline
6032364236763645056&1.14&nan&0.51&18.31&0.45&0.28&0.02&0.25&1.63\\
6032371177430994048&1.24&-0.40&0.44&16.68&0.4&0.15&0.39&0.07&1.73\\
6032371349229711616&0.96&0.40&0.45&17.26&0.42&0.01&0.16&0.42&1.55\\
6032372964137450240&1.19&1.64&0.47&17.9&0.51&0.32&0.12&0.04&1.59\\
6032408874375184896&1.06&nan&0.58&15.92&0.5&0.28&0.16&0.05&1.65\\
 \hline
\end{tabular}
\vspace*{2.5mm}
\caption{Metal-poor giant candidates found by Classifier-GP. (This table is available in its entirety in the online supplementary material)}
\label{table: Classifier_GP}
\end{table*}

\begin{figure*}
    \centering
    \subfloat[]{\includegraphics[scale=0.34]{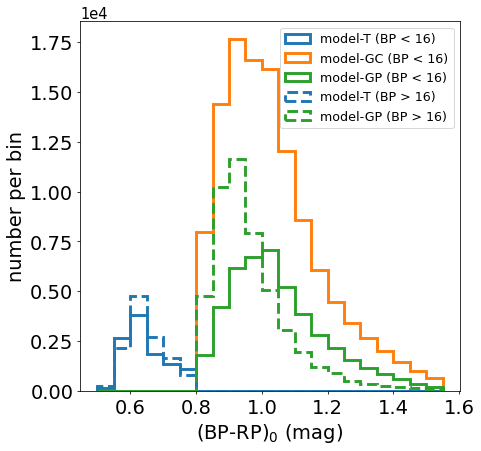}}
    \subfloat[]{\includegraphics[scale=0.34]{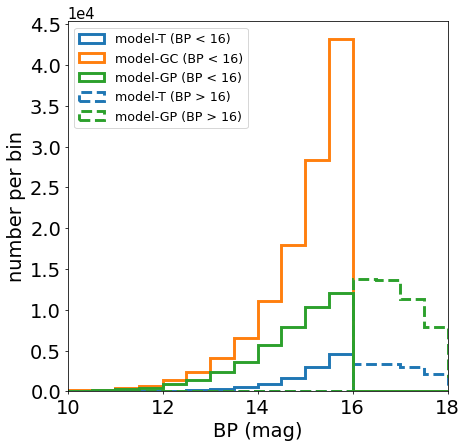}}
    \subfloat[]{\includegraphics[scale=0.34]{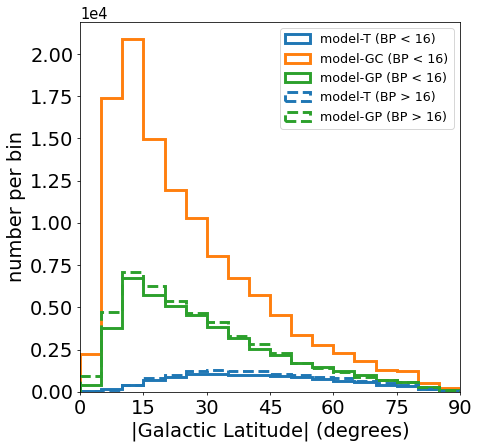}}\\
    \caption{$(BP-RP)_{0}$, $BP$, and $|b|$ distribution of the metal-poor candidates we found in Gaia DR3 by different classifiers. Note that the $|b|$ for giants is skewed to very low $|b|$ is because those are mostly towards the inner Galaxy (bulge/inner halo), as seen in Figure \ref{fig:distance_distribution}.}
    \label{fig: result_distribution}
\end{figure*}

\begin{figure*}
    \centering
    \includegraphics[scale=0.3]{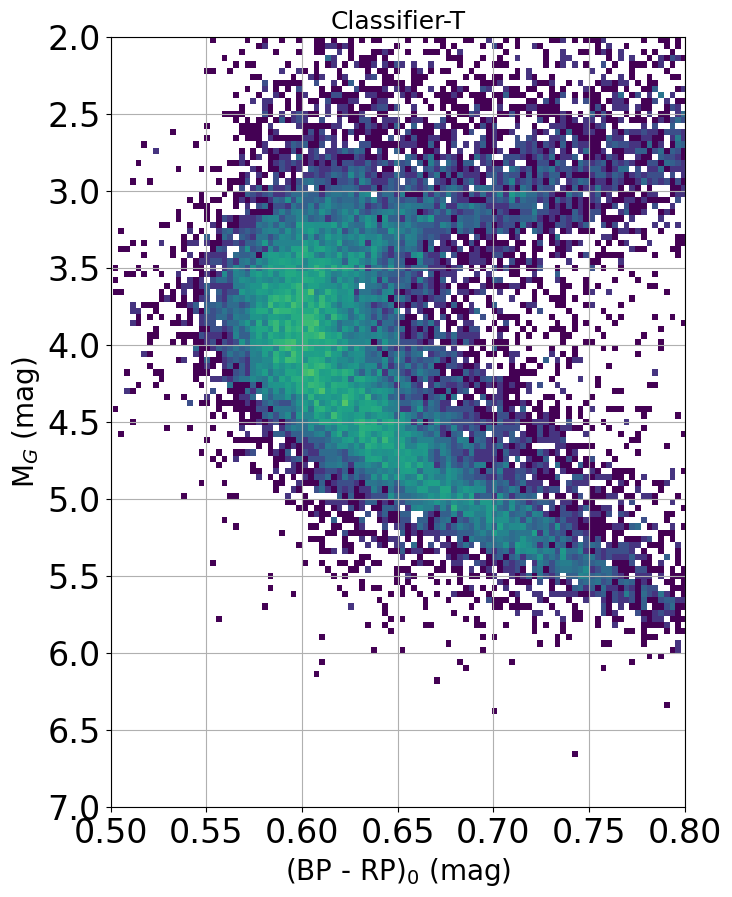}
    \includegraphics[scale=0.3]{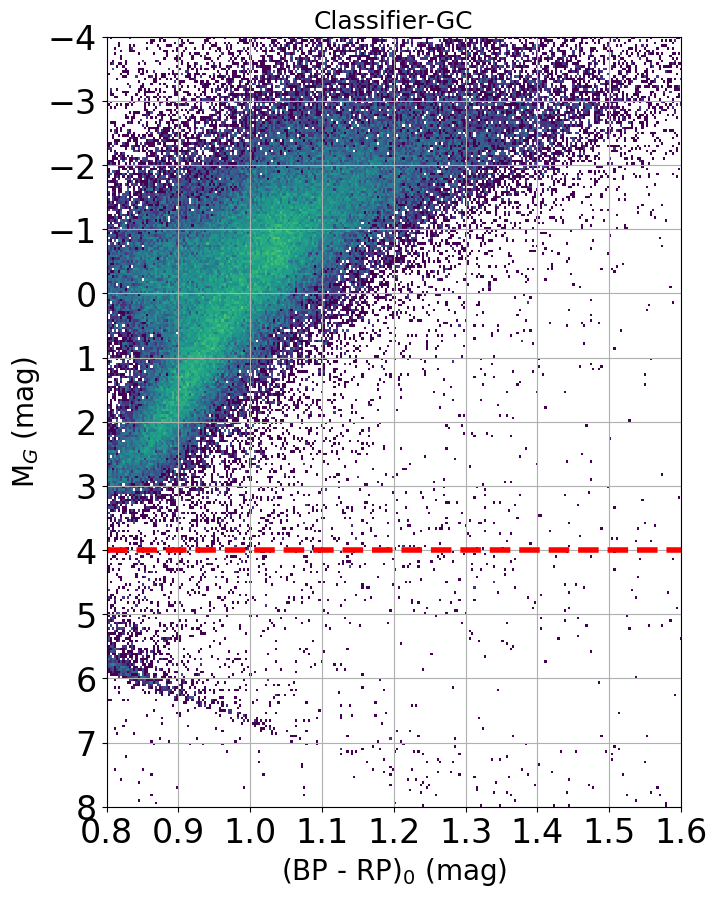}
    \includegraphics[scale=0.3]{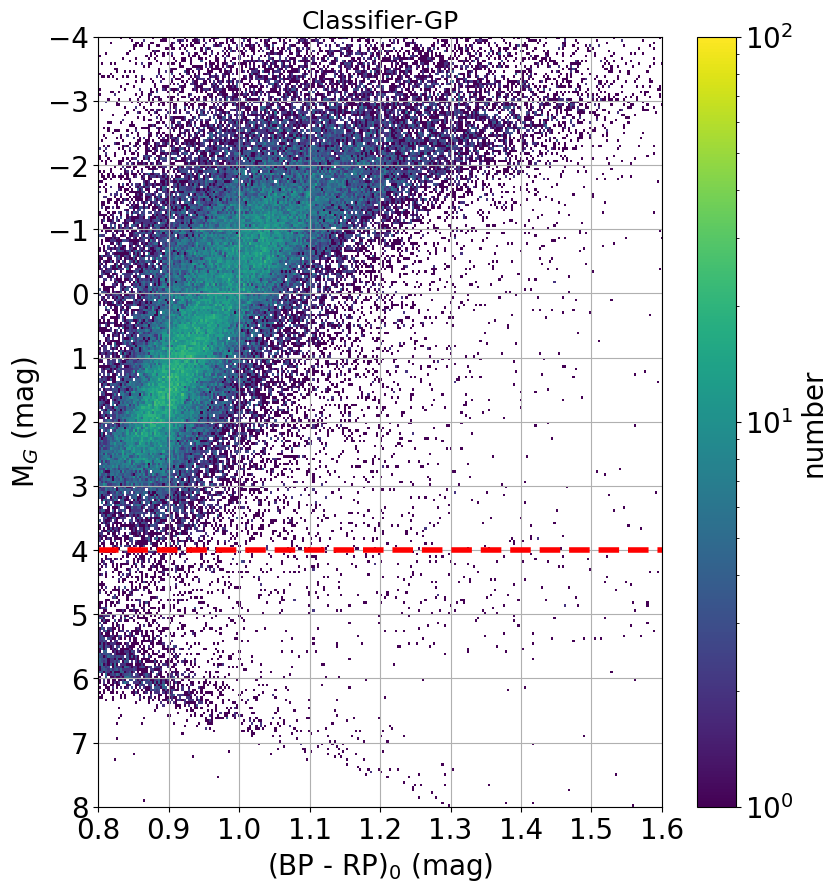}
    \caption{Colour-magnitude diagram of metal-poor stars we found in Gaia DR3 by different classifiers. The horizontal axis is dereddened colour $(BP-RP)_{0}$ and the vertical axis is the absolute $G$ band magnitude (for stars without any parallax-quality cut).}
    \label{fig:colour_magnitude_diagrams}
\end{figure*}

The colour-magnitude diagram for these candidate metal-poor stars, without any parallax quality cut, is shown in Figure \ref{fig:colour_magnitude_diagrams}. The left/middle/right panel shows the colour-magnitude diagram for the candidate metal-poor stars identified by Classifier-T/Classifier-GC/Classifier-GP. From these panels we confirm that, in our catalogues, the candidate metal-poor stars are dominated by turn-off stars and giant stars. However, there are a small number of dwarf stars present in the cooler regions of the main sequence, as shown in the middle and right panels ($M_{G} > 4$, below the red dashed line). These red dwarf stars may be wrongly classified as metal-poor stars, because there are almost no red dwarf stars in the training sets for Classifier-GP and -GC. Table \ref{table: dwarf_contamination} shows that red dwarfs only make up a very small fraction of the metal-poor stars found by Classifier-GC and -GP, i.e., 1.7\% for Classifier-GC, 0.7\% for Classifier-GP ($BP < 16$) and 6.5\% for Classifier-GP ($BP > 16$). Since the risk of contamination is higher, we include the absolute G band magnitude $M_{G}$ in our final catalogues if users would like to filter out any potential dwarf contamination.


\begin{table*}
\begin{tabular}{|p{4cm}|p{1.5cm}p{1.5cm}|p{1.5cm}p{1.5cm}|p{1.5cm}p{1.5cm}|}
\hline
Classifier and Brightness& \multicolumn{2}{c|}{Number of Stars} & \multicolumn{2}{c|}{Purity} & \multicolumn{2}{c|}{Completeness}\\
\hline
Classifier-T (BP < 16)&\multicolumn{2}{c|}{10995}&\multicolumn{2}{c|}{52\%}&\multicolumn{2}{c|}{32\%}\\
Classifier-T (BP > 16)&\multicolumn{2}{c|}{37763}&\multicolumn{2}{c|}{29\%}&\multicolumn{2}{c|}{8\%}\\
\hline
$M_{G}<4$ or $M_{G}>4$ & $M_{G}>4$ & $M_{G}<4$ & $M_{G}>4$ & $M_{G}<4$ & $M_{G}>4$ & $M_{G}<4$ \\
\hline
Classifier-GC (BP < 16)&1954&109493&27\%&45\%&56\%&94\%\\
Classifier-GP (BP < 16)&291&43514&50\%&76\%&10\%&66\%\\
Classifier-GP (BP > 16)&2542&38780&30\%&54\%&9\%&30\%\\
\hline
\end{tabular}
\vspace*{2.5mm}
\caption{The number, purity and completeness of metal-poor candidates we found by Classifier-T, Classifier-GC and Classifier-GP in different $M_{G}$ and BP ranges.}
\label{table: dwarf_contamination}
\end{table*}

\begin{table*}[h]
\begin{tabular}{|p{5cm}||p{2cm}|p{2cm}|p{2cm}|p{2cm}|p{2cm}|}
 \hline
 \multicolumn{6}{|c|}{Summary of the models} \\
 \hline
 model name &Classifier-T&Classifier-GC&Classifier-GP&Classifier-T&Classifier-GP\\
 \hline
 $(BP-RP)_{0}$   & $<$ 0.8 &$>$ 0.8&$>$ 0.8&$<$ 0.8&$>$ 0.8\\
 $BP$            & $<$ 16  &$<$ 16 &$<$ 16&$>$ 16&$>$ 16 \\
 Shannon Entropy Cutoff & nan & nan& nan&$<$ 0.8& nan\\
 Percentage of MP-stars  &0.06\%&0.28\%&0.28\%&0.06\%&0.28\%\\
 NPR of training set &1,000&40&386&1,000&386\\
 \ Test Purity        &47\%&47\%&74\%&40\%&53\%\\
 \ Test Completeness  &40\%&94\%&65\%&8\%&7\%\\
 \hline
 Expected Total Purity      &52\%&45\%&76\%&29\%&52\%\\
 Expected Total Completeness&32\%&93\%&66\%& 8\%&28\%\\
\# Candidates  &10,995&111,447&43,805&37,763&41,322\\
 \hline
\end{tabular}
\vspace*{2.5mm}
\caption{A summary table for Section 
\ref{Methodology} and \ref{Results}.}
\label{table: summary_cpn}
\end{table*}

\begin{figure}
    \centering
    \includegraphics[scale=0.7]{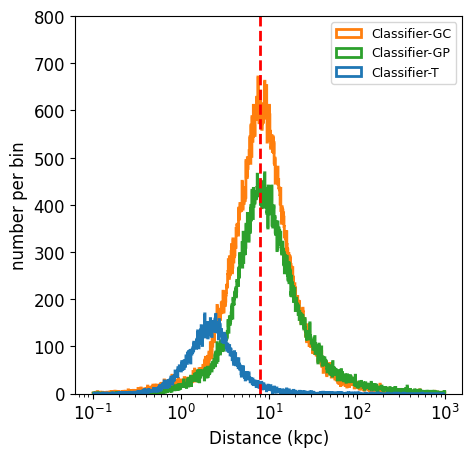}\\
    \caption{The distance distribution of the candidate metal-poor stars we found in Gaia DR3. The red dashed line in the left panel refers to the Galactic centre. Lines with different colour refer to the candidate metal-poor stars identified by different classifiers.}
    \label{fig:distance_distribution}
\end{figure}

\begin{figure*}
    \centering
    \includegraphics[scale=0.5]{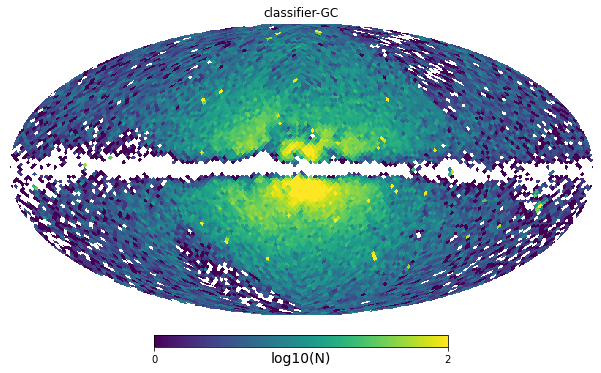}\\
    \includegraphics[scale=0.5]{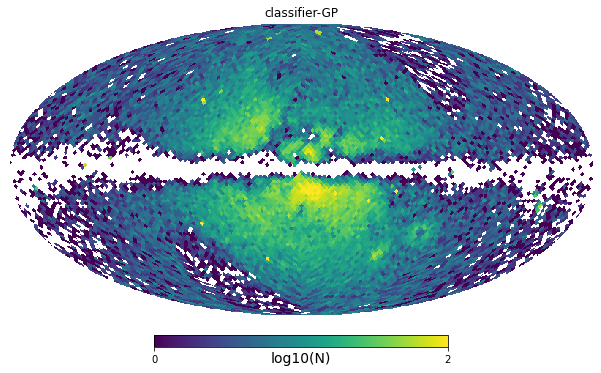}\\
    \includegraphics[scale=0.5]{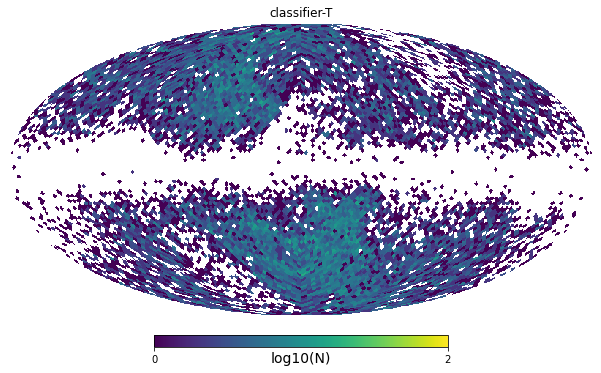}\\
    \caption{The galactic coordinate projections of the candidate metal-poor stars we found through Classifier-GC, Classifier-GP, Classifier-T. The area of healpix pixel is 3.36 deg$^2$}
    \label{fig:position_distribution}
\end{figure*}

The distance distributions and galactic coordinate projections of the candidates are shown in Figure \ref{fig:distance_distribution} and \ref{fig:position_distribution}. Figure \ref{fig:distance_distribution} shows the distance distributions of the candidate metal-poor stars. The distances are calculated by inverting the Gaia DR3 parallax. The distance to the Galactic centre is marked by the red dashed line \citep[$\sim$8 kpc from the Sun][]{bland2016galaxy}. The blue lines are the distribution of candidate turn-off metal-poor stars, and the orange and green lines are the distribution of candidate giant metal-poor stars. For the distance distribution, comparing to candidate metal-poor giant stars, the turn-off stars are located closer to the Sun, as expected given their lower luminosities. The giants are distributed around the Galactic centre. This result indicates that the Galactic centre contain a large amount of metal-poor stars, i.e., the Milky Way hosts an ancient, metal-poor, and centrally concentrated stellar population \citep[e.g.][]{rix2022poor}. Figure \ref{fig:position_distribution} shows the skymap of the candidate metal-poor stars we found in Gaia DR3. Because the dereddening process excludes almost all of the high $E(B-V)$ stars ($E(B-V)$ > 2), we do not obtain a lot of stars at low galactic latitude, as shown in figure \ref{fig:position_distribution}. Bulge stars and halo stars are the dominant stars for our sample. 

The bright spots in the Galactic coordinate projections are globular clusters \citep{harris2010new}. After testing, we found that comparing with Classifier-GC that includes many globular clusters with $-1.5 <$\rm[Fe/H]$< -1.0$, Classifier-GP excludes all of the globular clusters with average metallicity larger than $-$1.5 and most of the globular clusters with average metallicity within $-$2 to $-$1.5, but keeps all of the globular clusters with metallicity less than $-$2, which is a demonstration that Classifier-GP has relatively higher purity than Classifier-GC. Note that the galactic coordinate projections are also affected by the Gaia scanning law (see \citealt{de2022gaia}) and crowding issues for XP spectra in globular clusters. 

\section{Discussion}\label{Discussion}

In this work, according to Table \ref{table: dwarf_contamination}, we add up the numbers of metal-poor candidates found by Classifier-T, Classifier-GC ($BP < 16$ at all $M_G$), Classifier-GP ($BP > 16$ at all $M_G$) and obtained a total of 200,000 candidate metal-poor stars.
Weighting each subsample by its purity in Table~\ref{table: dwarf_contamination}, we expect the catalog contains 88,000 genuine metal-poor stars (overall purity of 44\%).

Though we only classify stars with $[\rm Fe/H] < -2$, we can estimate how many stars are $< -3$ or $< -4$. We assume the slope of the metallicity distribution  \citep{youakim2020pristine, chiti2021metal}, although there are also much more pessimistic slopes of the metal-poor tail of the metallicity distribution function by \citet{bonifacio2021topos}:

\begin{equation}
\log\frac{dN}{d\text{[Fe/H]}} = \gamma 
\label{eq:slope}
\end{equation}

$\gamma$ is 1 when $-$2.5 <$\rm[Fe/H]$< $-$2.0 and 1.5 in $-$4 <$\rm[Fe/H]$< $-$2.0. Based on this assumption, we can estimate the number of the actual metal-poor stars for these 188,000 candidate in each metallicity intervals. From $-$4 to $-$3.5, $-$3.5 to $-$3, $-$3 to $-$2.5, $-$2.5 to $-$2, the estimated number of actual metal-poor stars are 600, 2800, 17,000, 64,000 respectively.

\subsection{Comparing with other surveys}
Table \ref{table: comparing_1} shows our results compared to previous photometric selections. \citet{huang2022beyond} utilised SMSS DR2 and Gaia EDR3 photometry to estimate metallicity for 24 million stars. They obtained half a million very metal-poor ($\rm[Fe/H] < -2.0$) stars, and over 25,000 extremely metal-poor ($\rm[Fe/H] < -3.0$) stars. 48270 very metal-poor candidates in \citet{huang2022beyond} are also predicted to be very metal-poor by our Classifiers. \citet{chiti2021stellar} utilised SMSS DR2 photometry to derive photometric metallicities. They present more reliable metallicities of $\sim$280,000 stars with $-3.75 \le $$\rm[Fe/H]$$ \le -0.75$ down to $g = 17$. 18,640 of them are candidate metal-poor stars ($\rm[Fe/H] < -2$). After the validation by our training and testing set, we found their purity to be 49\%; and there are 9218 stars also predicted to be metal-poor by our Classifiers. Pristine survey does not publicly release their data, but according to \citet{starkenburg2017pristine} and \citet{youakim2020pristine}, Pristine has covered a sky area of $\sim$2500 $\text{deg}^{2}$, at the time of those papers. In each $\sim$$\text{deg}^{2}$ field, they find $\sim$7 stars that have $\rm[Fe/H] < -2.5$ down to magnitude of $V = 18$. The purity of Pristine to find stars with $\rm[Fe/H] < -2.5$ is 49\% \citep{aguado2019pristine}. The Best \& Brightest initiative selected over 11,000 candidate VMP ($\rm[Fe/H] < -2$) and EMP stars ($\rm[Fe/H] < -3$), with an overall purity of 30\% and 5\% respectively \citep{schlaufman2014best,placco2019r,limberg2021targeting}. Comparing with other surveys, our work increases the number of candidate metal-poor stars by about an order of magnitude, but with similar or higher purity. The comparison results are shown in Table \ref{table: comparing_1}. Recently, \citet{andrae2023robust} utilised \texttt{XGBoost} and XP spectra, together with 38 narrowband colours derived from XP spectra and broadband surveys (Gaia: G, BP, RP and CatWISE: W$_1$, W$_2$), to derive metallicity, Teff and logg for 175 million stars. They reduced the temperature-extinction degeneracy by introducing CatWISE W$_1$ and W$_2$, which extend to the infrared regions, into the model. The metallcity were derived using the \texttt{XGBoost} regression model and the true labels came from APOGEE, and augmented by a set of very metal-poor stars \citep{li2022four}. Because we both utilise \texttt{XGBoost} algorithm and deal with the same data set, it is worth comparing our results with them. The comparisons are shown in Table \ref{table: rene}. Table-1 and table-2 are two tables published by \citep{andrae2023robust}. In short, table-2 is a high accuracy subset of bright ($BP$ < 16) giant stars of table-1. Table \ref{table: rene} shows that, for giant candidates, Classifier-GP has higher purity and more candidates comparing with table-1 and table-2 (only including giant candidates with $BP$ < 16). The purity for turn-off stars of table-1 is only 20\%, while our Classifier-T has a higher purity of 29\% to 52\%. We suggest our models are better for finding metal-poor stars comparing with \citet{andrae2023robust} for the following reasons: (\RNum{1}) We have larger number of metal-poor stars, which provides the models a training set with greater diversity. (\RNum{2}) Their model is a regression model which is trying to fit the metallicity for all stars, especially for metal-rich stars. As a result, their model may not do as well for metal-poor stars, which are only a very small part of the whole. In contrast, our models are more specialized, and only focus on finding metal-poor stars. (\RNum{3}) Because we choose classification algorithm rather than regression, we can trade off completeness against purity. For stars that are difficult to classify, for example turn-off stars, we can sacrifice the completeness to the higher purity with NPR (see Fig. \ref{fig:npr_pc}) and SMOTE. (\RNum{4}) The Gaia XP spectra we utilised has been dereddened, which may makes our predictions more accurate, even without WISE photometry. Out of 148,000 very metal-poor candidates in \citet{andrae2023robust}, there are 65949 stars are found to be very metal-poor with our Classifiers. Overall, we suggest that researchers and observers utilise this work together with \citet{andrae2023robust} to decide what metal-poor candidates to follow up. 

\citet{zhang2023parameters} utilised a forward model to estimate stellar parameters ([Fe/H], $T_{\rm eff}$ and $logg$), revised distances and extinctions for 220 million stars with XP spectra. However, there is a trend that the metallicity derived by the forward model tend to be over-estimated at very-metal-poor end, which is even more biased than the metallicity derived by \citet{andrae2023robust}. We think this bias is caused by the imbalance of the numbers of the metal-poor and non-metal-poor stars in their training set.

\citet{martin2023pristine} used the spectroscopic and photometric information of 219 million stars from Gaia DR3 to calculate synthetic narrow-band $CaHK$ magnitudes sensitive to metallicity. $CaHK$ magnitudes mimic the observations of Pristine surveys. They derived the photometric metallicities for 30 million high signal-to-noise FGK stars. They identified 200,000 very metal-poor candidates and 8,000 extremely metal-poor candidates ($\rm [Fe/H]_{phot} < -2$ and $\rm [Fe/H]_{phot} < -3$ respectively). Because their data was released while this paper was already in review, we do not consider their results for our comparisons.

\begin{table}

\begin{tabular}{|p{3cm}||p{2.5cm}|p{1.2cm}|}
 \hline
 \multicolumn{3}{|c|}{Comparison to photometric selections} \\
 \hline
 Photometic Surveys & \# $\rm[Fe/H] < -2$ & Purity\\
 \hline
 SMSS(Turn-off)\citep{huang2022beyond}    &548,518    & 10\%\\
 SMSS(Giant)\citep{huang2022beyond}    &192,487    & 42\%\\
 SMSS(Turn-off)\citep{chiti2021stellar}      &522    & 46\%\\
 SMSS(Giant)\citep{chiti2021stellar}      &18,046   & 49\%\\
 Best\&Brightest\citep{placco2019r,limberg2021targeting}   &11,000 & 30\%\\
 Pristine \citep{starkenburg2017pristine, youakim2020pristine}  & 18,000*($\rm[Fe/H]<-2.5$)    & 49\%\\
 \hline
 Classifier-T       &48,758    & 29 - 52\%\\
 Classifier-GC      &111,447    & 45\%\\
 Classifier-GP      &85,127     & 52 - 76\%\\
 \hline
\end{tabular}
\vspace*{2.5mm}
\caption{Comparison to other photometric surveys. The purity mentioned above are obtained from the comparison with LAMOST DR7 and APOGEE DR17. Except for Pristine, for which
it is from \citet{aguado2019pristine}, and not for$\rm[Fe/H]$< $-$2 but $-$2.5.}  Note that, for Classifier-T or Classifier-GP, 29\% and 52\% are the purity for faint ($BP > 16$) stars, 52\% and 76\% are the purity for bright ($BP < 16$) stars.
\label{table: comparing_1}
\end{table}

\begin{table}
\begin{tabular}{|p{4cm}||p{2cm}|p{1.2cm}|}
 \hline
 \multicolumn{3}{|c|}{Comparison to \citet{andrae2023robust}} \\
 \hline
 Table/Model & \# $\rm[Fe/H] < -2$ & Purity\\
 \hline
 \citet{andrae2023robust} table-1 (Turn-off, $BP$ < 16) &24,000 & 23\%\\
  Classifier-T ($BP$ < 16) &10,995 & 52\%\\
 \hline
 \citet{andrae2023robust} table-1 (Giants, $BP$ < 16) &38,000 & 70\%\\
  Classifier-GP ($BP$ < 16) &43,805 & 76\%\\ 
  Classifier-GC ($BP$ < 16) &111,447 & 45\%\\ 
 \hline
 \citet{andrae2023robust} table-1 (Turn-off, $BP$ > 16) &51,000 & 20\%\\
  Classifier-T ($BP$ > 16, Shannon Entropy < 0.8) &37,763  & 29\%\\
 \hline
 \citet{andrae2023robust} table-1 (Giants, $BP$ > 16) &35,000 & 53\%\\
  Classifier-GP ($BP$ > 16) &41,322 & 52\%\\ 
 \hline
 \citet{andrae2023robust} table-2 & 18,000& 70\%\\
 \hline
\end{tabular}
\vspace*{2.5mm}
\caption{Comparison to \citet{andrae2023robust}. The purity mentioned above are obtained from the comparison with LAMOST DR7 and APOGEE DR17}
\label{table: rene}
\end{table}

\subsection{Validation with existing high-resolution spectra}
There are plenty of high-resolution follow-up observations to the candidate metal-poor stars that have been obtained by previous studies. We can utilise these confirmed metal-poor stars to evaluate the completeness of our \texttt{XGBoost} models. The results are shown in Table \ref{Golden}. In this table, we utilise 6 metal-poor halo stars data sets, 3 metal-poor bulge data sets, 1 metal-poor disk star and 1 carbon-enhanced metal-poor (CEMP; [C/Fe] > +0.7) data set to test our models. For each data set, we exclude stars of which dereddened colour $(BP-RP)_0 < 0.5$ and $E(B-V)$ > 2. Then we divide each data set into turn-off metal-poor stars ($(BP-RP)_0 < 0.8$) and giant metal-poor stars ($(BP-RP)_0 > 0.8$). The total number of these stars are shown in third and fourth columns. Finally, we utilise the Classifier-T, Classifier-GC and Classifier-GP to predict the metallicity of these turn-off and giant metal-poor stars respectively and get the corresponding completeness marked as completeness-T, completeness-GC and completeness-GP. This table shows that the completeness from these data sets is very close to the results from our test set, especially for the halo stars. We also test our classifiers on carbon enhanced metal-poor (CEMP) stars as shown in the last row of Table \ref{Golden}. As might be expected, the completeness of the classifiers on CEMP stars is not as high as other metal-poor stars, potentially because the enhanced carbon makes the metal-poor star spectra look more metal-rich. 

\section{Summary}\label{summary}
Metal-poor stars ($\rm[Fe/H] < -2$) record the chemical enrichment history, accretion events, and early stages of the Milky Way. However, they are rare and difficult to find. In this work, we train \texttt{XGBoost} models to identify metal-poor stars in Gaia DR3. The input to the models are the coefficients of normalized and dereddened XP spectra. The classifiers split the stars into different$\rm[Fe/H]$intervals of $-2.5 < \rm[Fe/H] < -2$, $-2 < \rm[Fe/H] < -$1.5, $-1.5 < \rm[Fe/H] < -1$, $-1 < \rm[Fe/H] < +1$. Because of the extreme imbalance between positive and negative samples, we randomly exclude some negative samples and utilise the SMOTE algorithm to over-sample the training sets and, then, utilise them to train the models. Finally, we get three classifiers, Classifier-T, Classifier-GC, and Classifier-GP and utilise them to identify the metal-poor turn-off and giant stars in Gaia DR3 with XP spectra. We present the histogram of the testing result and the completeness/purity distributions for these models in Figure \ref{fig:metallicity_distribution} and Figure \ref{fig: CP_dis}.

In total, we obtained 200,000 metal-poor candidates with overall purity 44\%. This number of metal-poor candidates is around an order of magnitude larger than previous work (e.g., Best \& Brightest, SkyMapper, and Pristine), which has similar or even better purity.  

We make the full catalog available in the supplementary online material. Table \ref{table: Classifier_T}, \ref{table: Classifier_GC}, \ref{table: Classifier_GP}

\begin{table*}
\setlength{\arrayrulewidth}{0.5mm}
\setlength{\tabcolsep}{20pt}
\renewcommand{\arraystretch}{1.5}
\begin{tabular}{|p{2.0cm}|p{0.4cm}|p{0.2cm}|p{0.2cm}|p{1.7cm}|p{1.7cm}|p{1.7cm}|}
\hline
Reference&Region or Type&MP-Giants&MP-turn-off&Completeness-GC&Completeness-GP&Completeness-T\\
\hline
\citet{abohalima2018jinabase}         & Halo   & 266 &115 &  98.9\% & 82.3\% & 54.8\% \\
\citet{li2022four}& Halo   & 152 & 79 &  92.8\% & 71.1\% & 55.7\% \\
\citet{ roederer2014search}    & Halo   &  68 & 111 & 100.0\% & 77\% & 56\% \\
\citet{jacobson2015high}    & Halo   &  106 &  1 & 100.0\% & 83\% & 0\%     \\
\citet{cohen2013normal}       & Halo   &  47 &  26 &  94\% & 64\% & 42\%     \\
\citet{cayrel2004first}& Halo   & 25 & 0 &  100\% & 48\% & NAN \\
\citet{sestito2022pristine}     & Bulge  &   8 &  5 & 100.0\% & 37.5\% & 40.0\% \\
\citet{ howes2015extremely}       & Bulge  &  21 &  0 & 100.0\% & 71\% & NAN     \\
\citet{howes2016embla}      & Bulge  &   9 &  0 & 100.0\% & 56\% & NAN     \\
\citet{schlaufman2018ultra}      & Disk  &   0 &  1 & NAN & NAN & 100\%     \\
\citet{yoon2016observational} & CEMP   &   106 &  34 &     84\%  &  55\% & 41\%\\
\hline
\end{tabular}
\vspace*{2.5mm}
\label{table: gs}
\caption{\label{Golden} Prediction results for metal-poor stars that are confirmed by High-resolution spectra.}
\end{table*}

\section*{ACKNOWLEDGEMENTS}
We thank Yang Huang, Xiaowei Ou and Anirudh Chiti for helpful discussions.  Y.Y. and A.P.J. acknowledge support from the U.S. National Science Foundation (NSF) grant AST 2206264. G.L. acknowledges FAPESP (procs. 2021/10429-0 and 2022/07301-5). This research benefited from the 2022 Gaia DR3 Chicago Sprint hosted by the Kavli Institute for Cosmological Physics. We acknowledge the University of Chicago’s Research Computing Center for their support of this work.

This work has made utilise of data from the European Space Agency (ESA) mission Gaia (\url{https://www.cosmos.esa. int/gaia}), the Large Sky Area Multi-Object Fiber Spectroscopic Telescope and Apache Point Observatory Galactic Evolution Experiment. Mission Gaia is processed by the Gaia Data Processing and Analysis Consortium (DPAC, \url{https://www.cosmos.esa.int/web/gaia/ dpac/consortium}). Funding for the DPAC has been provided by national institutions, in particular the institutions participating in the Gaia Multilateral Agreement. The Large Sky Area Multi-Object Fiber Spectroscopic Telescope (LAMOST) is a National Major Scientific Project built by the Chinese Academy of Sciences. Funding for the project has been provided by the National Development and Reform Commission. LAMOST is operated and managed by the National Astronomical Observatories, Chinese Academy of Sciences. Apache Point Observatory Galactic Evolution Experiment (APOGEE) is one of the programs in the Sloan Digital Sky Survey \RNum{3}. Funding for the creation and distribution of the SDSS Archive has been provided by the Alfred P. Sloan Foundation, the Participating Institutions, the National Aeronautics and Space Administration, the National Science Foundation, the U.S. Department of Energy, the Japanese Monbukagakusho, and the Max Planck Society. The SDSS Web site is \url{http://www.sdss.org/}. The Participating Institutions are The University of Chicago, Fermilab, the Institute for Advanced Study, the Japan Participation Group, The Johns Hopkins University, the Max-Planck-Institute for Astronomy (MPIA), the Max-Planck-Institute for Astrophysics (MPA), New Mexico State University, Princeton University, the United States Naval Observatory, and the University of Washington.

This research  makes utilise of public auxiliary data provided by ESA/Gaia/DPAC/CU5 and prepared by Carine Babusiaux.
This paper made utilise of the Whole Sky Database (wsdb) created by Sergey 
Koposov and maintained at the Institute of Astronomy, Cambridge with financial support  from the Science \& Technology Facilities Council (STFC) and the European Research Council (ERC).

This research made utilise of Astropy\footnote{\url{http://www.astropy.org}} a community-developed core Python package for Astronomy \citep{robitaille2013astropy}. Other software utilised includes matplotlib \citep{hunter2007matplotlib}, numpy \citep{harris2020array}, scikit-learn \citep{pedregosa2011scikit} and scipy \citep{virtanen2020scipy}.

\section*{DATA AVAILABILITY}
Our result, i.e., all contents in Table \ref{table: Classifier_T} ,\ref{table: Classifier_GC} ,\ref{table: Classifier_GP} , can be aquired at: \url{https://doi.org/10.5281/zenodo.8360958}

\bibliographystyle{mnras}
\bibliography{bibliography}

\appendix
\section{Prediction uncertainty}\label{Appendix_a}

\begin{table*}
\setlength{\arrayrulewidth}{0.2mm}
\setlength{\tabcolsep}{20pt}
\renewcommand{\arraystretch}{1.5}
\begin{tabular}{|p{4.5cm}|p{1.5cm}|p{1.0cm}|p{1.5cm}|p{1.0cm}|}
\hline
 &\multicolumn{2}{c|}{[Fe/H] < $-$1.5}&\multicolumn{2}{c|}{[Fe/H] < $-$1.0}\\
\hline
 &Completeness & Purity &Completeness & Purity\\
\hline
Classifier-T, BP < 16              & 36\% & 79\% & 61\% & 90\%\\
Classifier-T, BP > 16               & 34\% & 75\% & 55\% & 91\%\\
Classifier-GC, BP < 16, $M_{G} > 4$ & 67\% & 86\% & 77\% & 93\%\\
Classifier-GC, BP < 16, $M_{G} < 4$ & 17\% & 74\% & 17\% & 80\%\\
Classifier-GP, BP < 16, $M_{G} > 4$ & 71\% & 87\% & 83\% & 93\%\\
Classifier-GP, BP < 16, $M_{G} < 4$ & 27\% & 66\% & 31\% & 72\%\\
Classifier-GP, BP > 16, $M_{G} > 4$ & 48\% & 75\% & 58\% & 88\%\\
Classifier-GP, BP > 16, $M_{G} < 4$ & 31\% & 75\% & 21\% & 80\%\\
\hline
\end{tabular}
\vspace*{2.5mm}
\caption{Completeness and purity of the classifiers for stars with$\rm[Fe/H]$< -1.5 or -1.0}
\label{table: P1P2}
\end{table*}

Shannon Entropy is an indicator of the prediction uncertainty, which is defined as:
\begin{equation}
\text{Shannon\;entropy} = - \sum_{i = 0}^{3} P_{i}\log_2(P_{i})
\label{eq:shannon entropy}
\end{equation}

\begin{figure*}
    \includegraphics[scale=0.3]{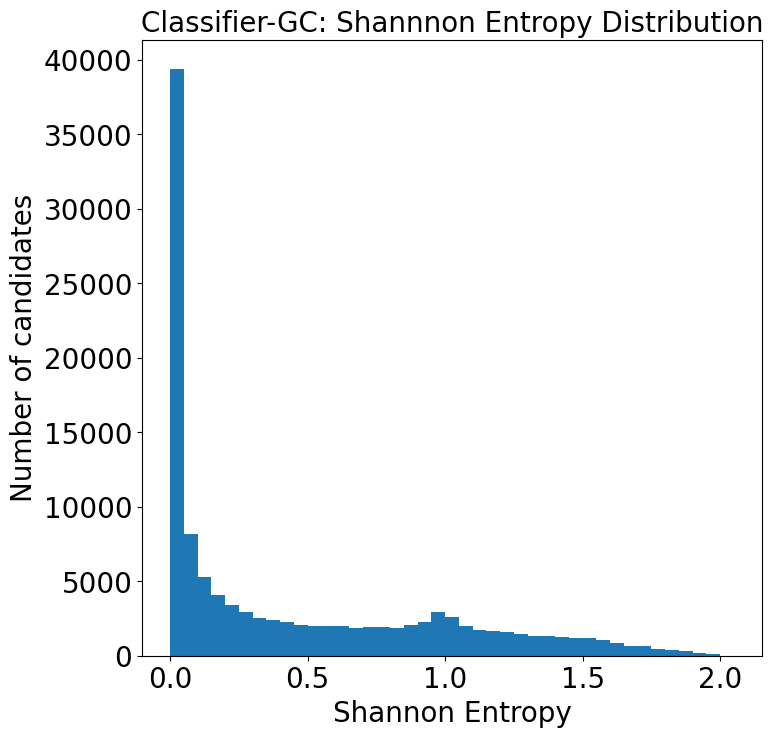}
    \includegraphics[scale=0.3]{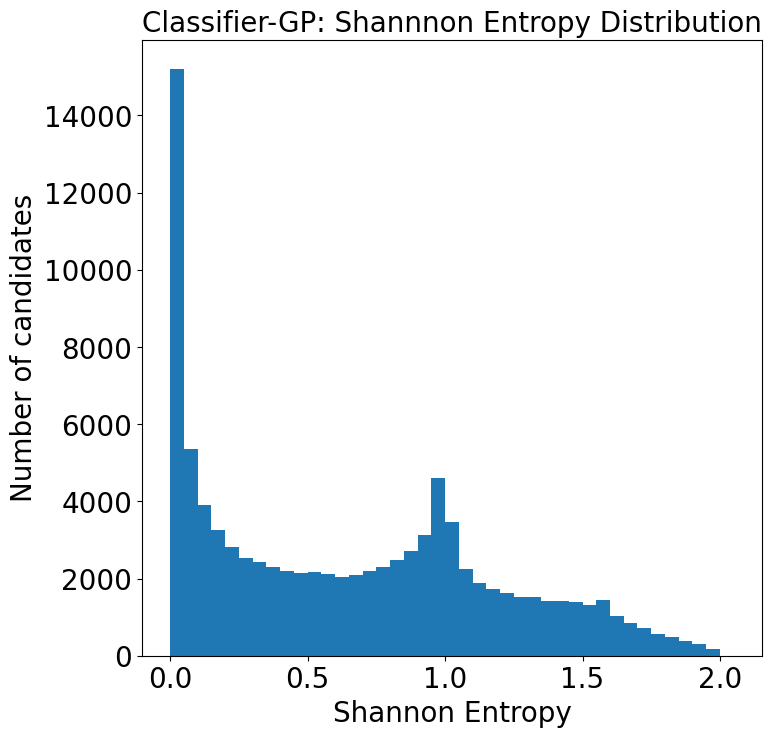}
    \includegraphics[scale=0.3]{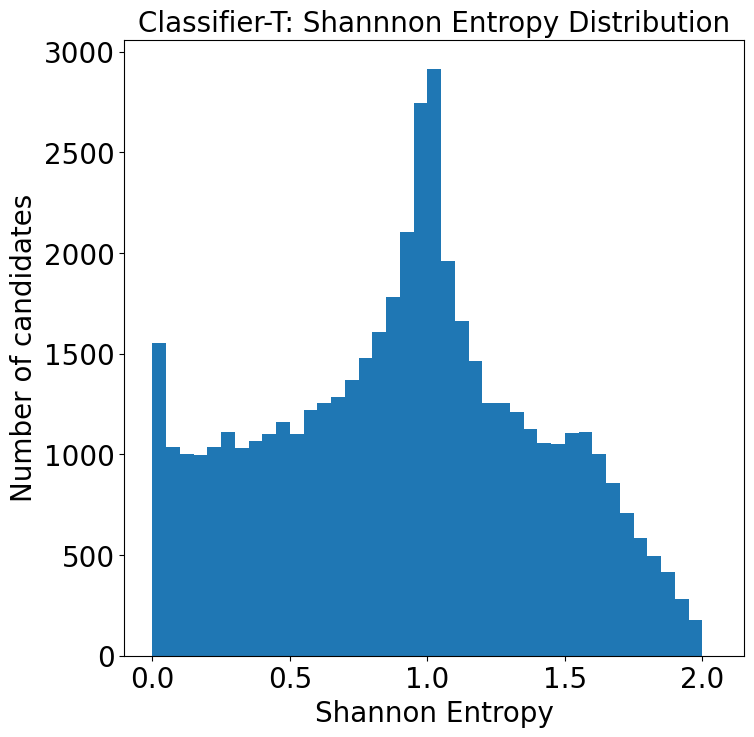}
    \caption{Shannon Entropy distributions of the metal-poor candidates found by different Classifiers.}
    \label{fig: shannon entropy distribution}
\end{figure*}

\begin{figure*}
    \includegraphics[scale=0.25]{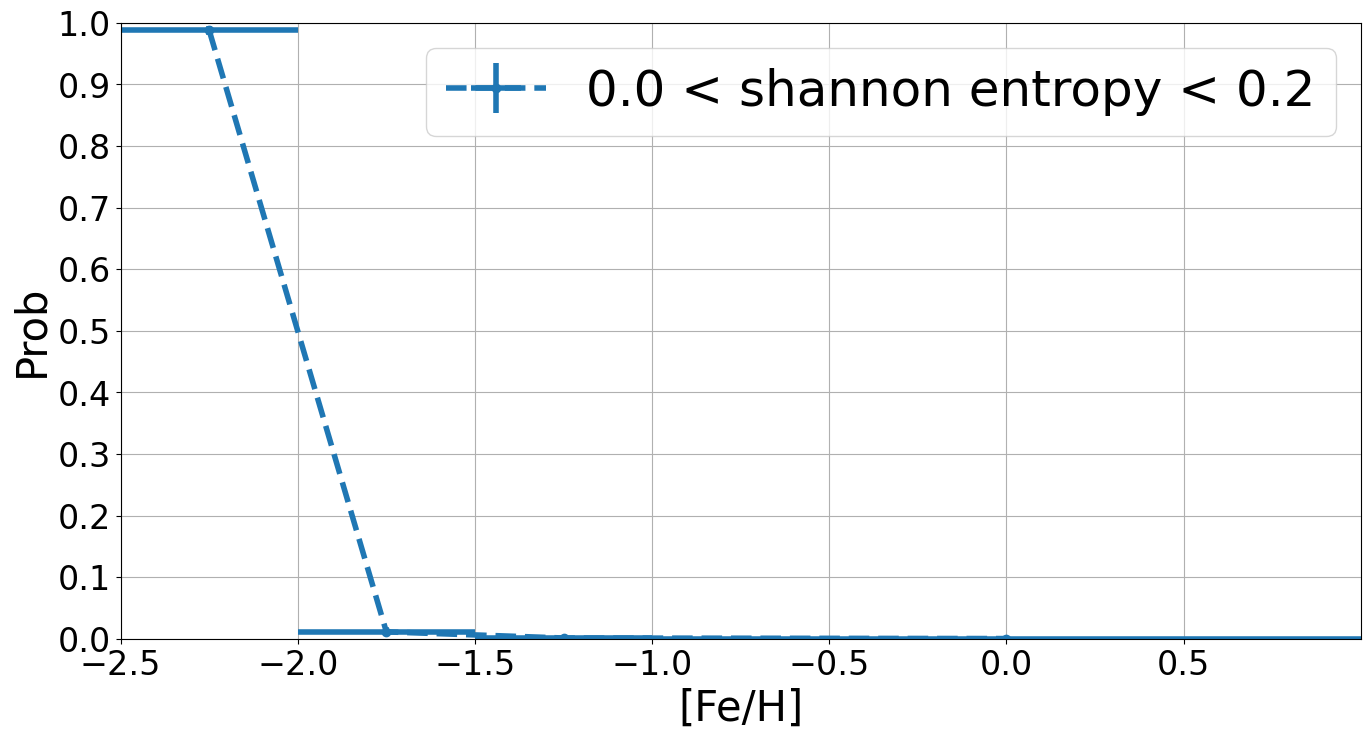}
    \includegraphics[scale=0.25]{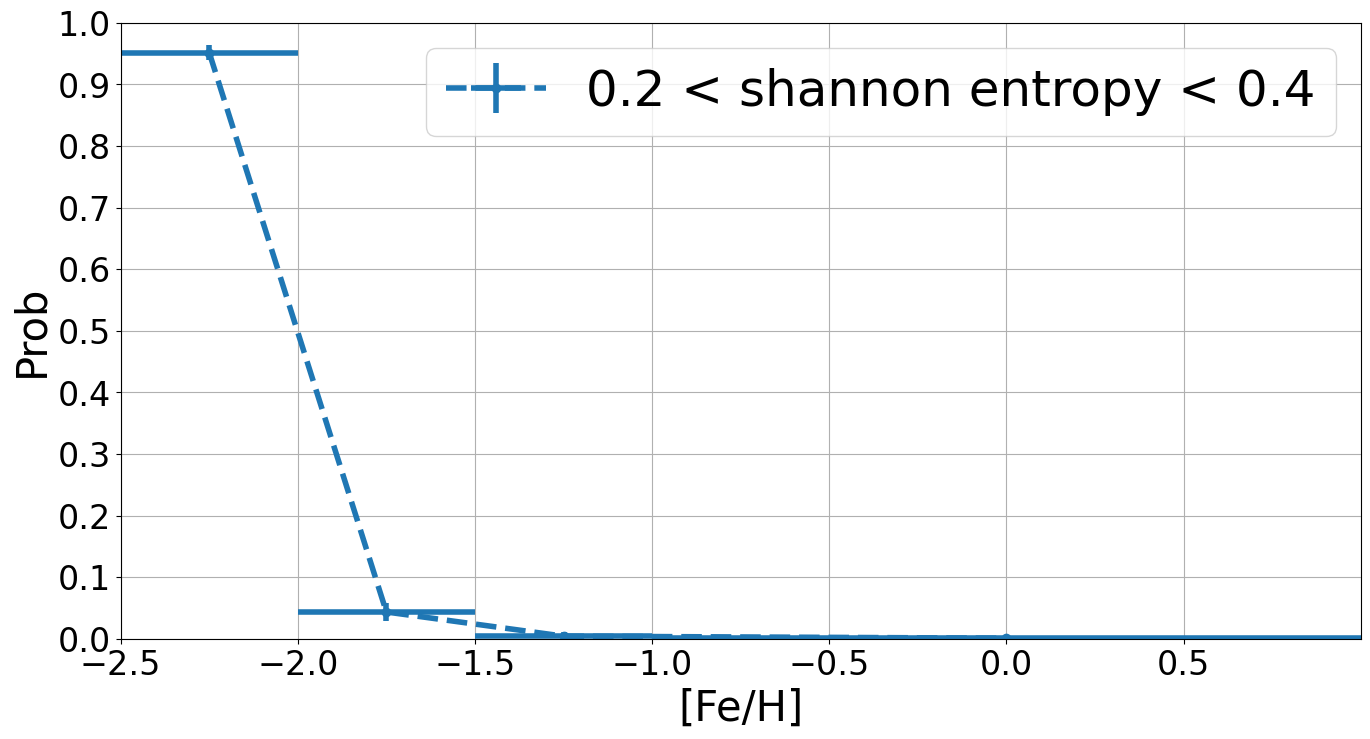}
    \includegraphics[scale=0.25]{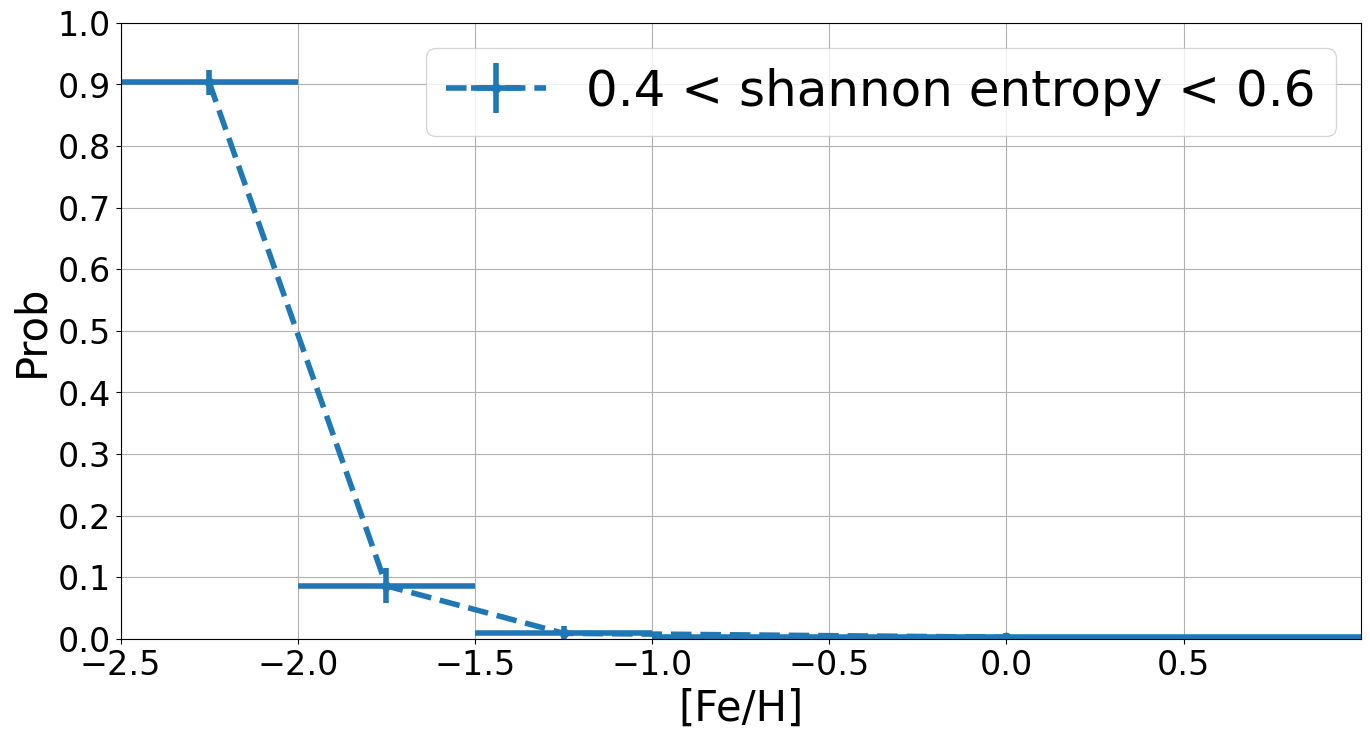}
    \includegraphics[scale=0.25]{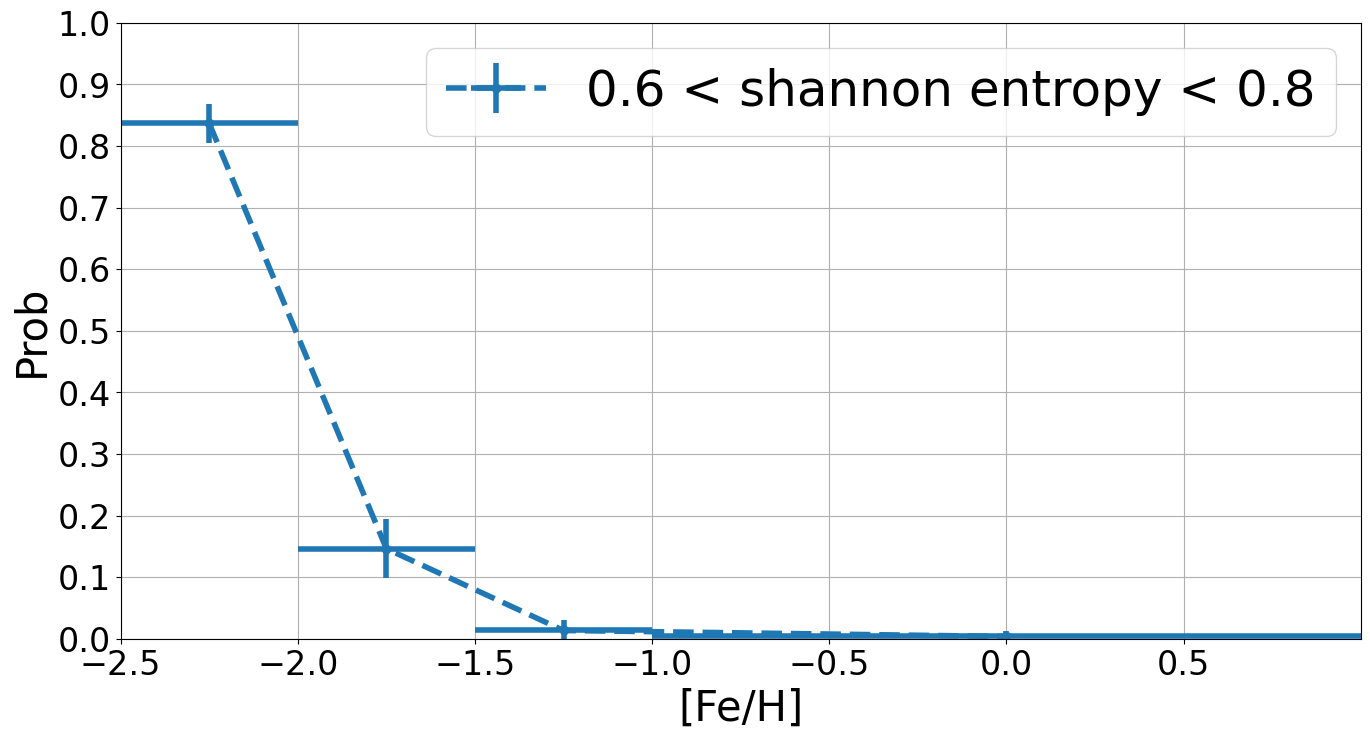}
    \includegraphics[scale=0.25]{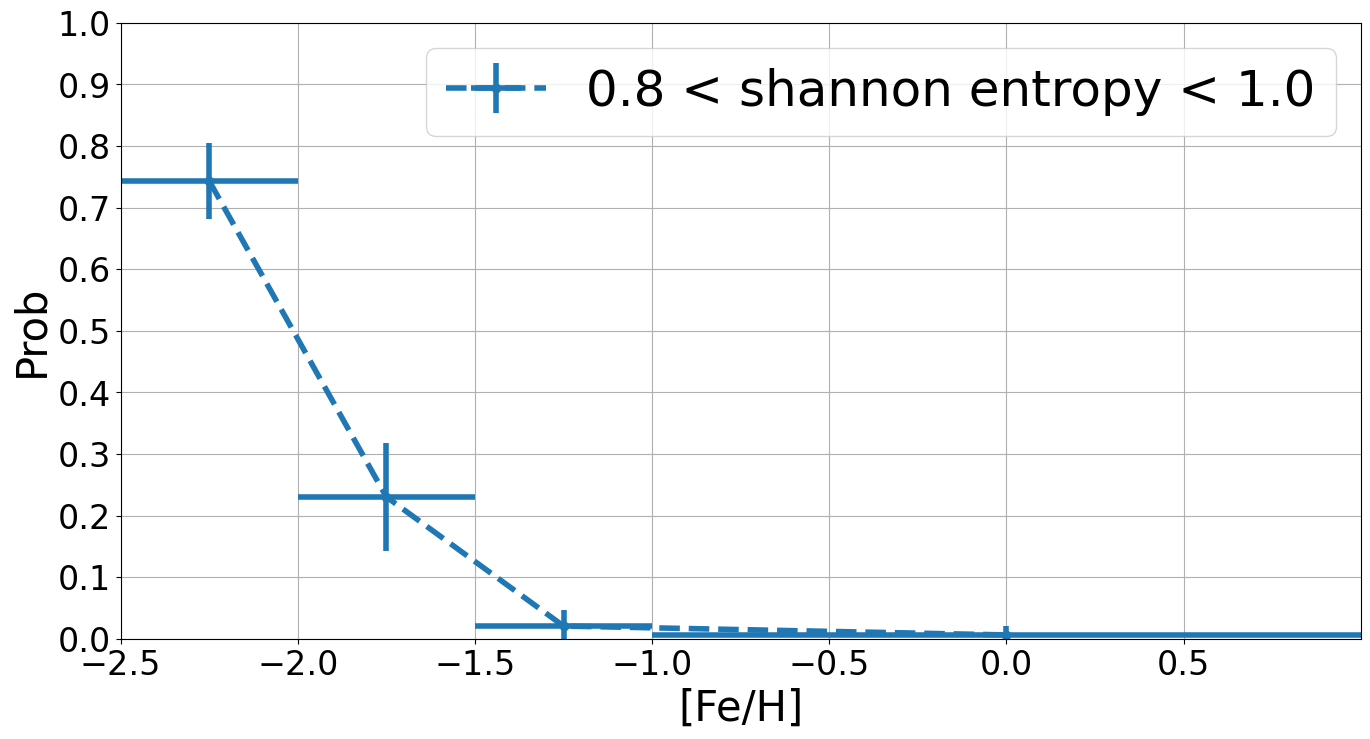}
    \includegraphics[scale=0.25]{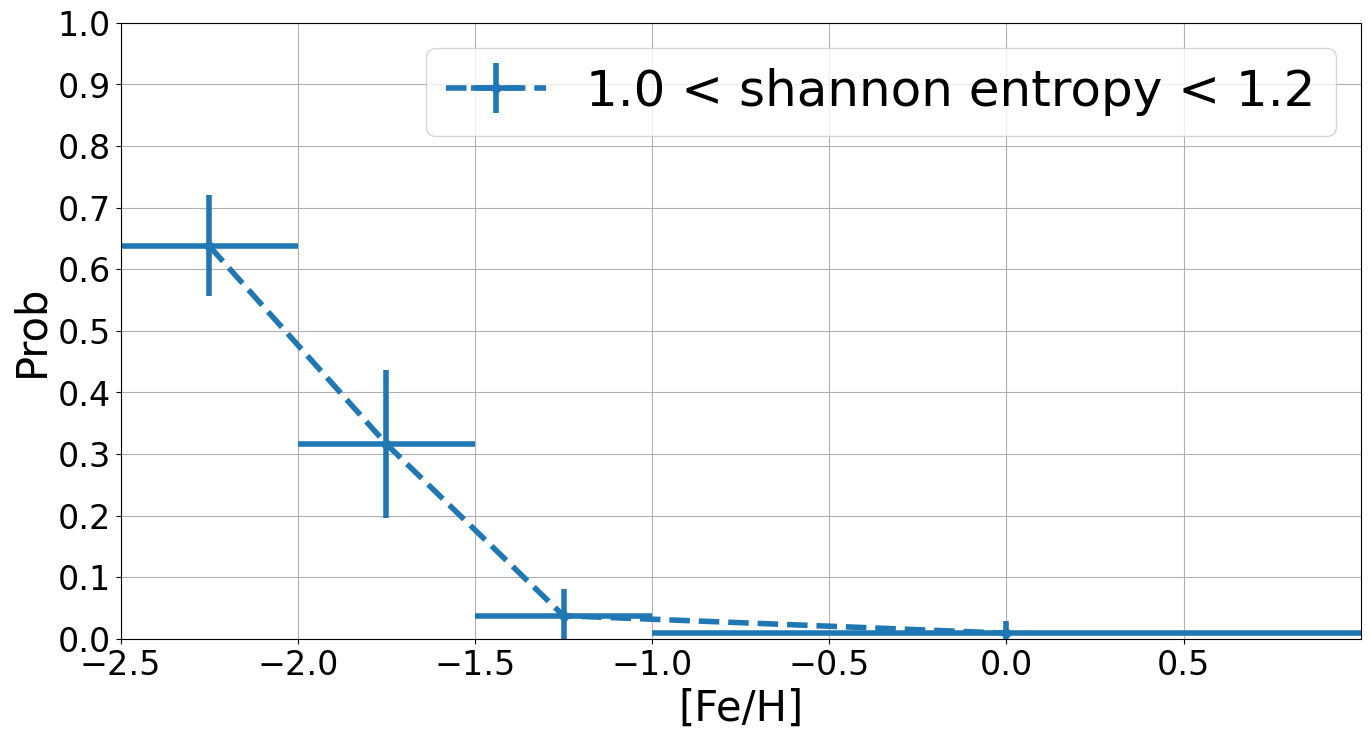}
    \includegraphics[scale=0.25]{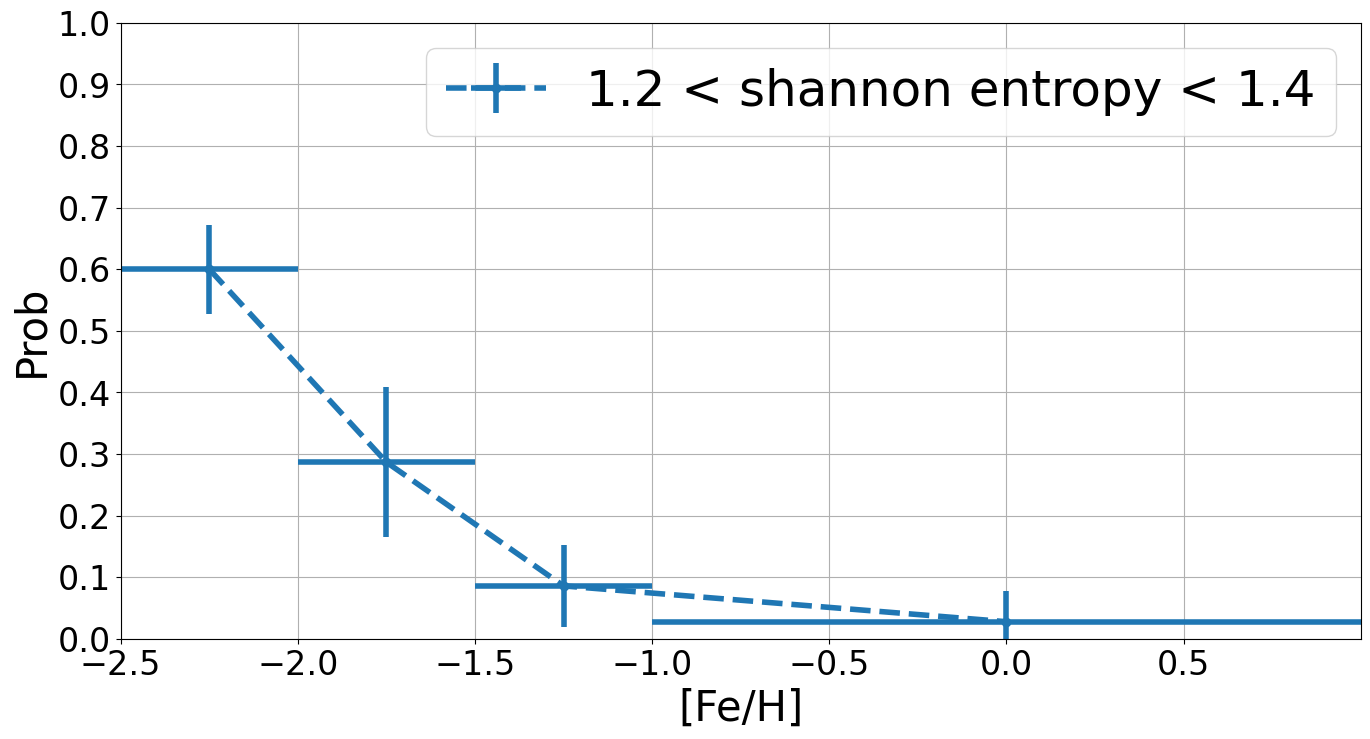}
    \includegraphics[scale=0.25]{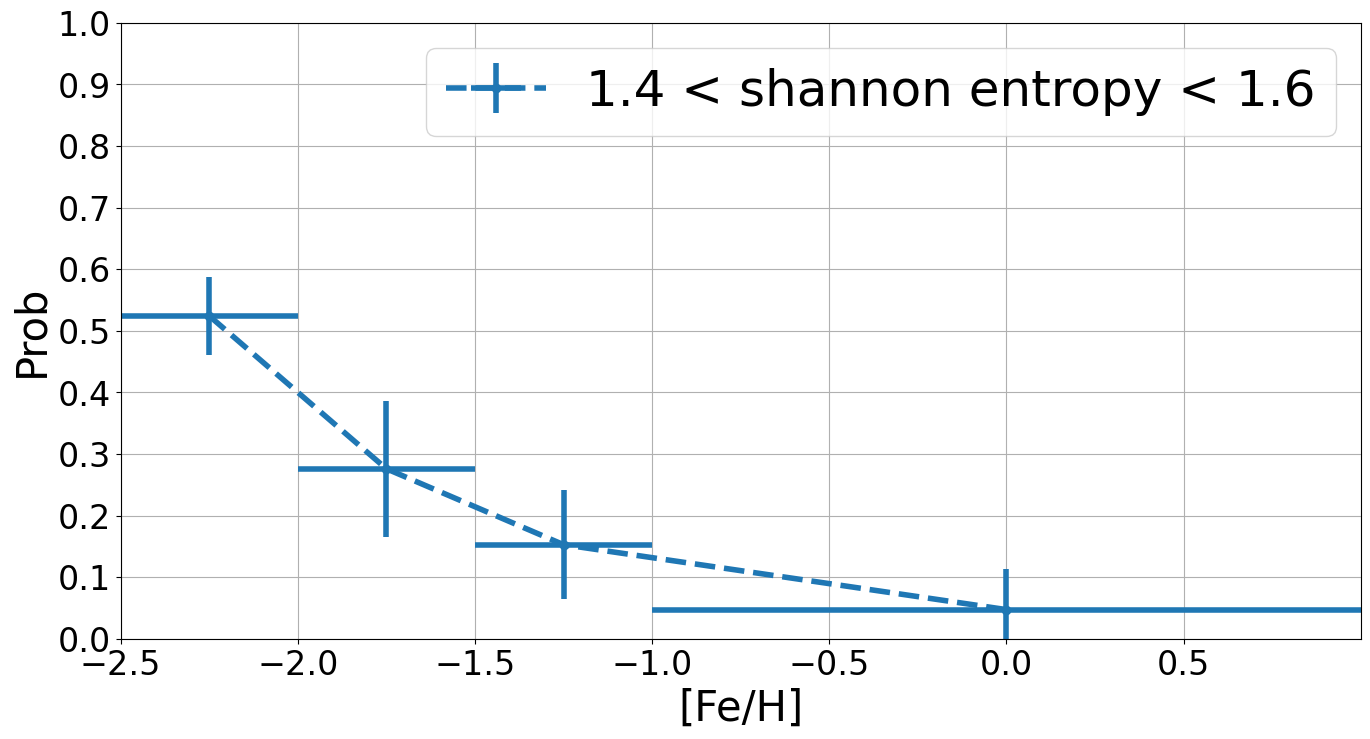}
    \includegraphics[scale=0.25]{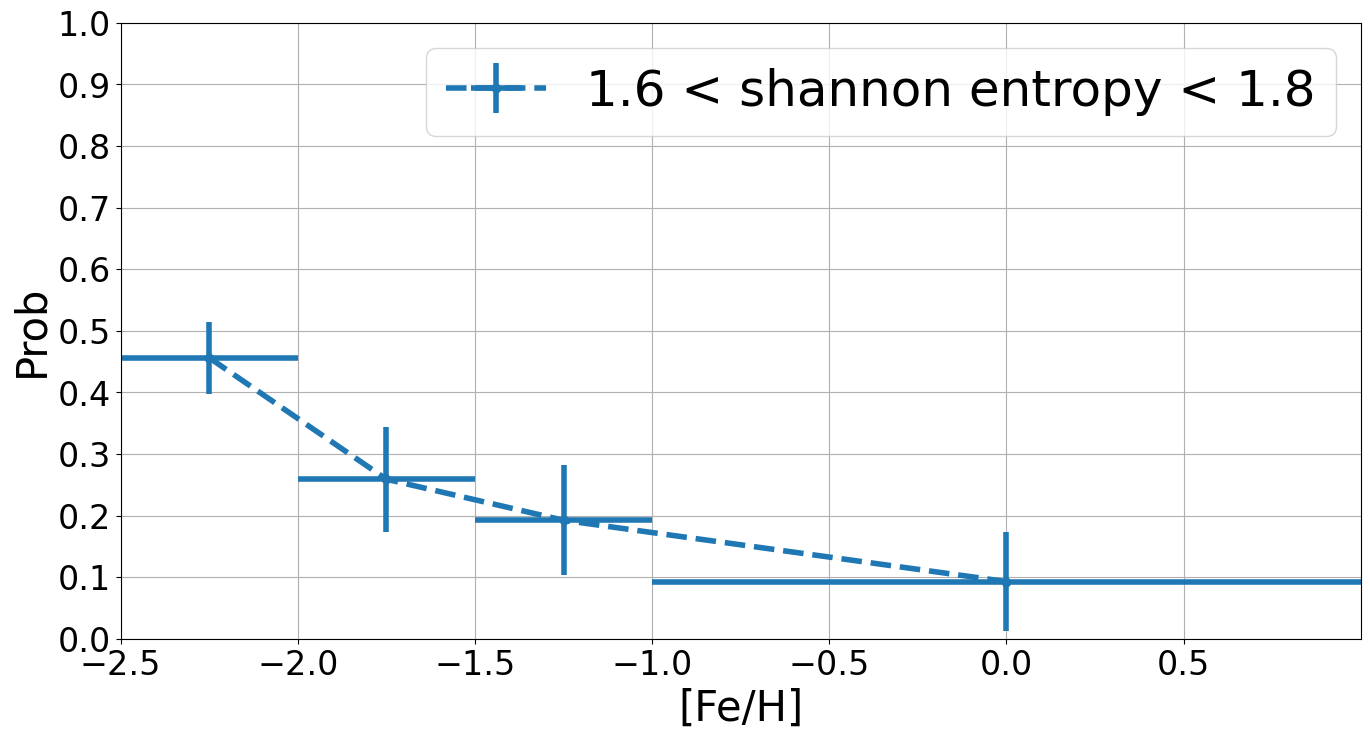}
    \includegraphics[scale=0.25]{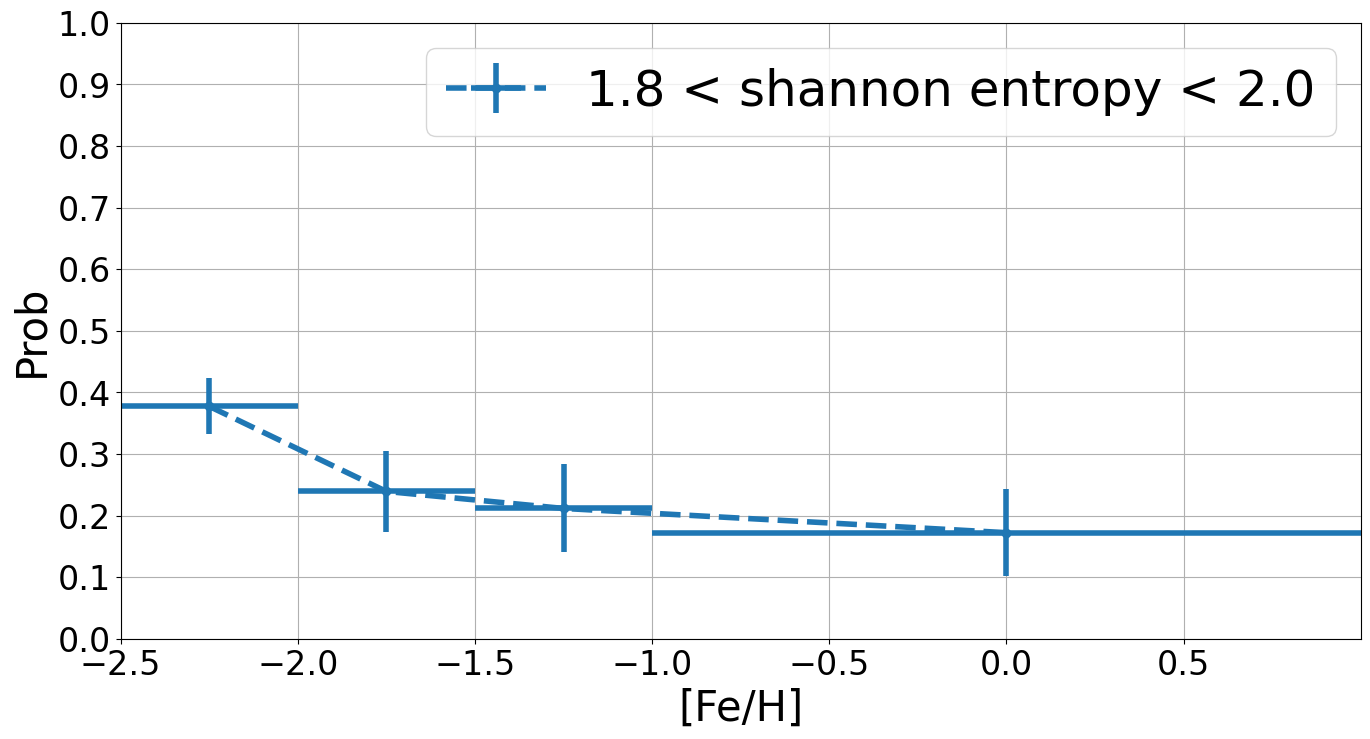}
    \caption{Distributions of mean $P_{0}, P_{1}, P_{2}, P_{3}$ with 1 $\sigma$ error bars of our three catalogues in different Shannon Entropy intervals.}
    \label{fig: se_vs_prob}
\end{figure*}

\begin{figure}
    \includegraphics[scale=0.5]{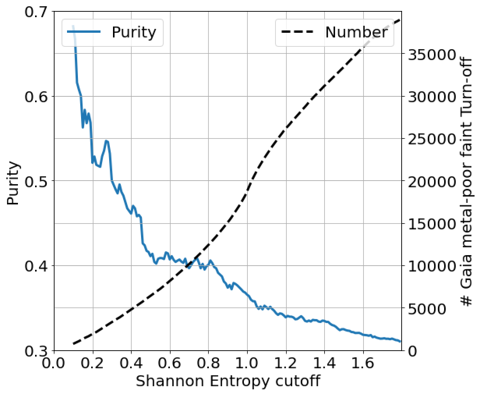}
    \caption{Number and purity of the remaining metal-poor faint ($BP$ > 16) turn-off candidates as a function of Shannon Entropy threshold}
    \label{fig: shannon_entropy}
\end{figure}

Shannon Entropy is an indicator of prediction uncertainty, which can be utilised to filter the metal-poor candidates with high prediction uncertainty and increase the purity of catalogues. According to the definition \ref{eq:shannon entropy}, Shannon Entropy increases as the probabilities become evenly distributed and decrease as they become skewed distributed. Thus, higher Shannon Entropy typically indicates greater prediction uncertainty. In this project, since we utilise multi-classification algorithm, each star in our catalogues is assigned four probabilities $P_{0}, P_{1}, P_{2}$, and $P_{3}$ (summing to 1) that correspond to the probabilities of the star belonging to four metallicity intervals: $\rm[Fe/H] < -2$, $-2 < \rm[Fe/H] < -1.5$, $-1.5 < \rm[Fe/H] < -1$, $-1 < \rm[Fe/H] < -1$. By comparing Figure \ref{fig: shannon entropy distribution} and Figure \ref{fig: se_vs_prob}, we see that most of our candidates have Shannon Entropy smaller than 1.5, which indicates that most of the candidates have $P_{0}$ greater than 0.5, in other words, most of the candidates have low prediction uncertainty. Even so, we can still increase the purity of our catalogues by excluding the stars with high Shannon Entropy (high prediction uncertainty). For example, as shown in Figure \ref{fig: shannon_entropy}, by excluding the candidates with Shannon Entropy $>$ 0.8, we can get a faint ($BP$ > 16) metal-poor turn-off star catalogue with purity > 40\%.

Additionally, our catalogues are also useful for the science goals requiring stars with $\rm[Fe/H] < -1.5$ or $\rm[Fe/H] < -1.0$. As shown in Table \ref{table: P1P2}, our Classifiers can also accurately and completely identify stars with $\rm[Fe/H] < -1.5$ or $\rm[Fe/H] < -1.0$. Comparing with finding stars with $\rm[Fe/H] < -2.0$, finding stars with $\rm[Fe/H] < -1.5$ or $\rm[Fe/H] < -1.0$ is an easier task because there are many more positive samples in our training and testing sets for these tasks.

\begin{figure*}
    \includegraphics[scale=0.3]{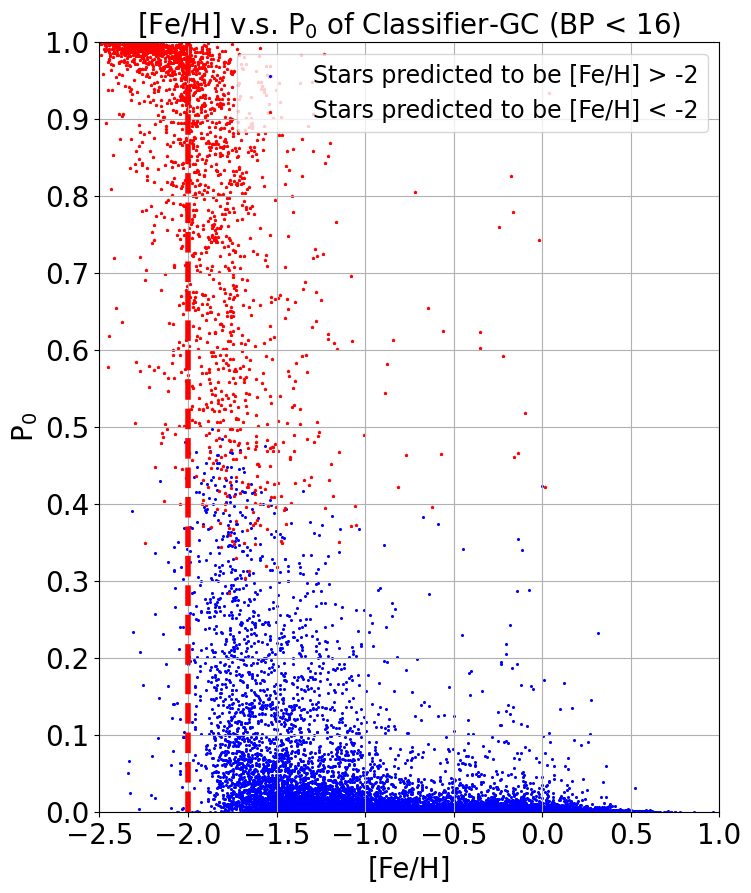}
    \includegraphics[scale=0.3]{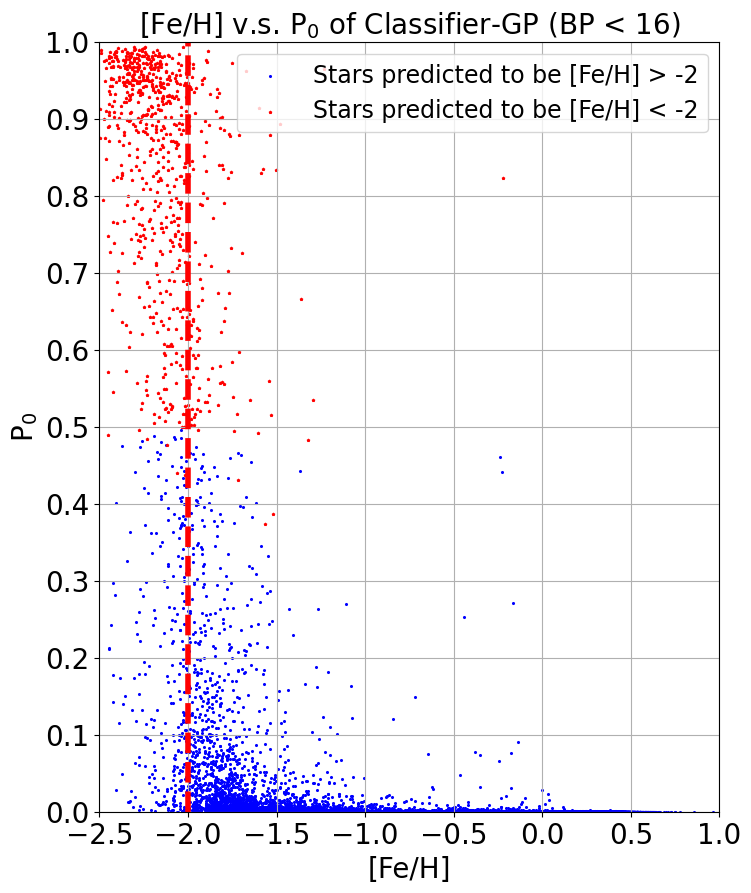}
    \includegraphics[scale=0.3]{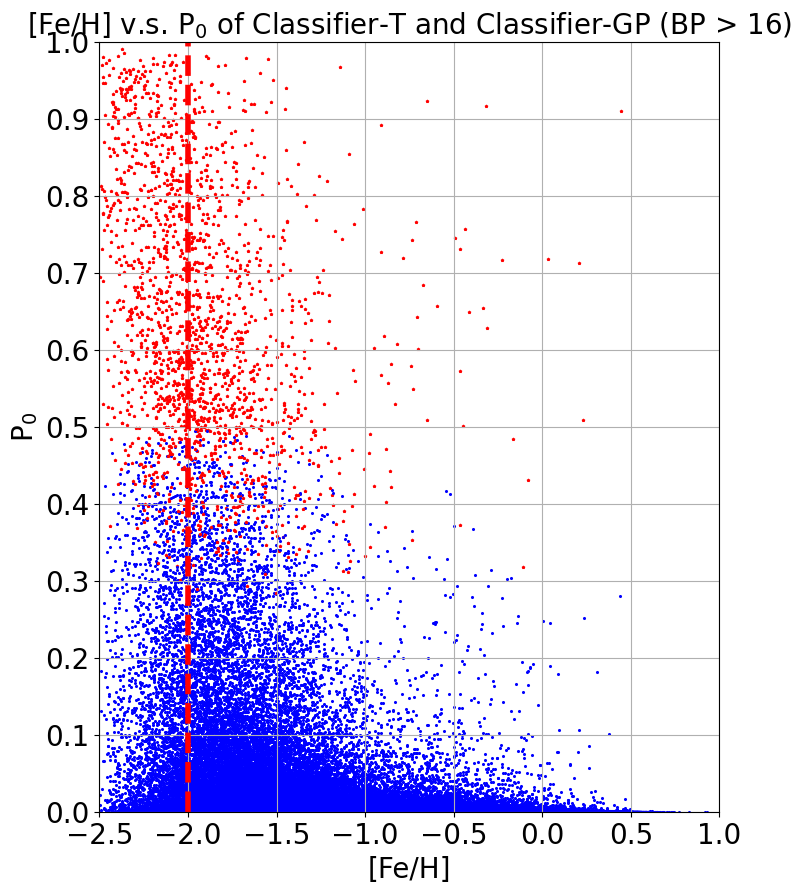}
    \caption{$P_{0}$ as a function of [Fe/H]}
    \label{fig: feh_vs_prob}
\end{figure*}

It is important to see probability-distribution situations for the stars with $\rm[Fe/H]$ close to -2 because, as shown in Figure \ref{fig:metallicity_distribution}, there are a lot of stars with $\rm[Fe/H]$ close to -2 in our catalogues. Figure \ref{fig: feh_vs_prob} shows $\rm[Fe/H]$ v.s. $P_{0}$ of testing sets. The left and middle panels are for Classifier-GC and Classifier-GP on bright stars (BP < 16), and the right panel is for Classifier-GP on faint stars (BP > 16) and Classifier-T. In these panels, red points refer to the stars predicted to be metal-poor, and blue points refer to those predicted to be non-metal-poor. The left and middle panels of Figure \ref{fig: feh_vs_prob} show that, as the increase of $\rm[Fe/H]$ from -2.5 to -1.5, $P_{0}$ sharply decreases from 1 to nearly 0, which indicates that the $P_{0}$ of Classifier-GC and Classifier-GP are sensitive to the metallicity variance (for bright stars). Additionally, there are a lot of blue points (false-negative samples) with $\rm[Fe/H] < -2$ in the right panel of Figure \ref{fig: feh_vs_prob}, because we sacrificed the completeness of turn-off stars and faint giant stars to get higher purity, as discussed in Section \ref{Methodology}. Fortunately, however, most of the red points in the right panels are still metal-poor, which is a sign of high purity.

\section{Extinction correction of the BP/RP coefficients} 
\label{sec:extinction}

Since the BP and RP coefficients of the star with extinction will differ from the coefficients of the same star without extinction, here we  try directly correct the BP/RP coefficients for extinction effects. 
The extinction model we assume is the following. If $C$ is the BP/RP coefficient vector of a star without extinction (the coefficient vector is normalized by the first coefficient). We assume that the effects of extinction can be described as 
\begin{align}
    C_{extincted}- C= (\alpha + \beta  C_0)  E_{B-V} + \gamma E_{B-V}^2 \label{eqn:ext}
\end{align}

where $\alpha,\gamma$ are vectors with the same number of elements as the length of the coefficient vector, and $\beta$ is a matrix.
The rationale behind this parametrisation is that the first term in the right hand side of the equation is providing linear changes of coefficients with extinction and the extinction coefficients can differ for stars with different spectra (this is essentially a Taylor expansion). The final term allows some non-linearity of the coefficients with extinction (but without dependence on the coefficients themselves).

To fit for the coefficients we take the APOGEE DR17 catalog with {\it Gaia} BP/RP coefficients. For each star  with extinction $E_{B-V}>0.05$ and BP/RP coefficients $C$ we find a nearest neighbor in the space of $T_{eff}, \log g, \rm[Fe/H]$ but with $E_{B-V}<0.05$. This provides us with the estimated unextincted BP/RP vector for that star. We then fit the relation (Eq.~\ref{eqn:ext}) between unextincted and extincted coefficients using regularized linear regression (as implemented in class {\tt LassoCV} in {\tt sklearn} package). We have found that the extinction coefficients mostly dependent on first few BP/RP coefficients (as those determine the broad spectral shape), thus we force the matrix $\beta$ to only have first 10 non-zero rows.
We provide the best fit  $\alpha, \beta,\gamma$ for BP/RP in supplementary materials.

The figure~\ref{fig:extinction} demonstrates the effect of the exctinction corrections. The top rows shows the differences between the coefficients of extincted vs non-extincted stars vs extinction. We can clearly see that several coefficients show strong dependence on $E_{B-V}$ as expected. The bottom panels shows what happens after correcting the coefficients. We can see that the trends with extinction mostly disappeared. 

In this extinction correction we rely on the \citet{schlegel1998maps} 2-D maps thus we essentially make an assumption that all of the stars are behind the dust layer. When this assumption is broken we expect that our corrections will not be appropriate.

\begin{figure*}
    \centering
    \includegraphics{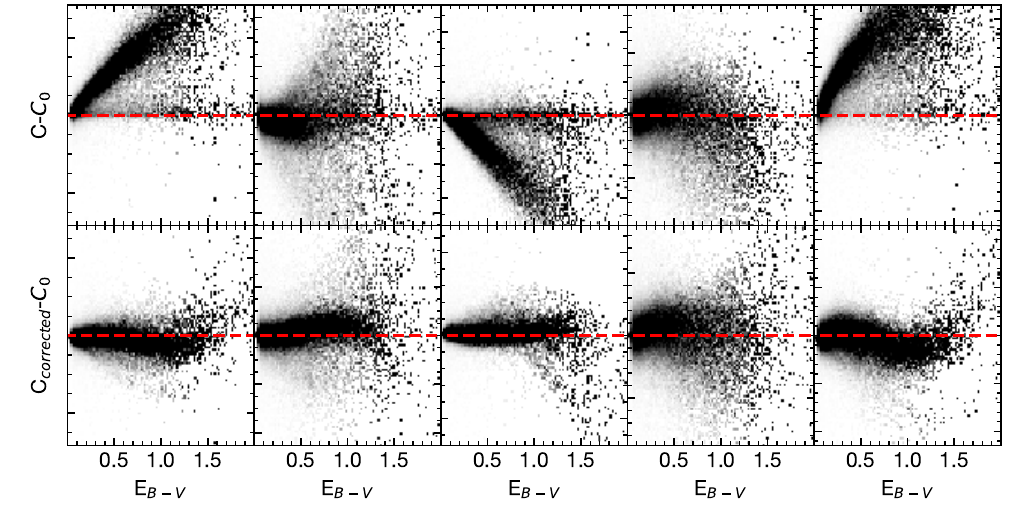}
    \caption{Effect of extinction correction. The top panel show the difference between BP coefficients between stars with zero extinction and stars with same stellar atmospheric parameters, but significant extinction vs value of extinction. The bottom panel shows the same but after applying the best fit extinction correction from Section~\ref{sec:extinction}. The red line shows where zero is.}
    \label{fig:extinction}
\end{figure*}

\section{WSDB ARCHIVE QUERIES}\label{queries}
Following ADQL queries were utilised to cross-match LAMOST and APOGEE to Gaia DR3 source id with XP spectra on WSDB:

\textbf{Query utilised for APOGEE:}
\begin{verbatim}
select 
bp_chi_squared,
rp_chi_squared,
bp_degrees_of_freedom,
rp_degrees_of_freedom,
sfd_ebv,gaiaedr3_phot_g_mean_mag,
source_id,fe_h,alpha_m,
logg,teff,ra,dec,
gaiaedr3_parallax, 
gaiaedr3_parallax_error, {COEFFS}

 from apogee_dr17.allstar as a,
 gaia_dr3.xp_continuous_mean_spectrum as s 
 
where
s.source_id=a.gaiaedr3_source_id
and bp_chi_squared < 1.5*bp_degrees_of_freedom
and rp_chi_squared < 1.5*rp_degrees_of_freedom
\end{verbatim}

\textbf{Query utilised for LAMOST:}
\begin{verbatim}
with 
x as(select gaia_source_id, feh, teff, logg, 
rank() over 
(partition by gaia_source_id order by snrr desc) 
from lamost_dr7.lrs_stellar),

y as (select gaia_source_id::bigint 
as sid,feh,teff,logg from x where rank=1),

z as (select feh,teff,logg, 
(select dr3_source_id 
from gaia_edr3.dr2_neighbourhood as g
where g.dr2_source_id=sid 
order by angular_distance asc limit 1)
as source_id from y where sid>0)

select 
bp_chi_squared, rp_chi_squared,
bp_degrees_of_freedom, rp_degrees_of_freedom,
feh, ebv, phot_g_mean_mag,
g.source_id, teff,
logg, g.ra, g.dec, {COEFFS}

from z as a, 
gaia_dr3.xp_continuous_mean_spectrum as s,
gaia_dr3.gaia_source as g 
where g.source_id= a.source_id 
and s.source_id=g.source_id

\end{verbatim}

\bsp	
\label{lastpage}
\end{document}